\documentclass[a4paper,fleqn,usenatbib]{mnras}  

\usepackage[T1]{fontenc}
\usepackage{ae,aecompl}

\usepackage{graphicx}	
\usepackage{amsmath}	
\usepackage{amssymb}	%
\usepackage{deluxetable}

\bibpunct{(}{)}{;}{a}{}{,}
\usepackage{hyperref}
\hypersetup{
    colorlinks,
    citecolor=blue,
    filecolor=blue,
    linkcolor=blue,
    urlcolor=blue
}

\title[Tidal Stripping with SIDM]{Enhanced tidal stripping of satellites in the galactic halo from dark matter self-interactions}

\author[G. A. Dooley et al.]{\parbox{17cm}{
Gregory A. Dooley$^{1}$\thanks{e-mail: gdooley@mit.edu},
Annika H.G. Peter$^{2,3}$,
Mark Vogelsberger$^{1}$,
Jes\'us Zavala$^{4}$\thanks{Marie Curie Fellow},
and Anna Frebel$^{1}$
}\vspace{0.3cm}\\
$^{1}$Department of Physics, Kavli Institute for Astrophysics and Space Research, Massachusetts Institute of Technology, \\ \ \ 77 Massachusetts Avenue, Cambridge, MA 02139, USA\\
$^{2}$CCAPP and Department of Physics, The Ohio State University, 191 W. Woodruff Ave., Columbus, OH 43210, USA \\
$^{3}$Department of Astronomy, The Ohio State University, 140 W. 18th Ave., Columbus OH 43210, USA\\
$^{4}$Dark Cosmology Centre, Niels Bohr Institute, University of Copenhagen, Juliane Maries Vej 30, DK-2100 Copenhagen, Denmark
}

\date{Accepted XXX. Received YYY; in original form ZZZ}

\pubyear{2016}

\newcommand{\vi}{\textbf{v-i }}
\newcommand{\vd}{\textbf{v-d }}

\begin{document}
\label{firstpage}
\pagerange{\pageref{firstpage}--\pageref{lastpage}}
\maketitle

\begin{abstract}  
We investigate the effects of self-interacting dark matter (SIDM) on the tidal stripping and evaporation of satellite galaxies in a Milky Way-like host. We use a suite of five zoom-in, dark-matter-only simulations, two with velocity-independent SIDM cross-sections, two with velocity-dependent SIDM cross-sections, and one cold dark matter (CDM) simulation for comparison. After carefully assigning stellar mass to satellites at infall, we find that stars are stripped at a higher rate in SIDM than in CDM. In contrast, the total bound dark matter mass-loss rate is minimally affected, with subhalo evaporation having negligible effects on satellites for viable SIDM models. Centrally located stars in SIDM haloes disperse out to larger radii as cores grow. Consequently, the half-light radius of satellites increases, stars become more vulnerable to tidal stripping, and the stellar mass function is suppressed. We find that the ratio of core radius to tidal radius accurately predicts the relative strength of enhanced SIDM stellar stripping. Velocity-independent SIDM models show a modest increase in the stellar stripping effect with satellite mass, whereas velocity-dependent SIDM models show a large increase in this effect towards lower masses, making observations of ultrafaint dwarfs prime targets for distinguishing between and constraining SIDM models. Due to small cores in the largest satellites of velocity-dependent SIDM, no identifiable imprint is left on the all-sky properties of the stellar halo. While our results focus on SIDM, the main physical mechanism of enhanced tidal stripping of stars apply similarly to satellites with cores formed via other means.

\end{abstract}


\section{Introduction}
The cold dark matter (CDM) paradigm is part of a simple model that successfully describes the Universe on large scales. Assuming that gravity alone acts on dark matter particles, the theoretical CDM framework has been shown to adequately explain halo mass functions, correlation functions, and shapes of large galaxies, clusters, and filaments (for a review, see \citealt{Frenk12}). On smaller scales, however, tensions with observations still arise when using CDM to predict properties of substructure and dwarf galaxies. For instance, whereas simulations predict high central density cusps in dark matter haloes \citep{Navarro97, Bullock01, Wechsler02}, observations of low surface brightness galaxies \citep{deBlok97,deBlok01,Kuzio11}, low-mass spiral galaxies, \citep{Gentile04, Simon05, Oh11, Castignani12, Adams14}, and Milky Way dwarf spheroidal galaxies \citep{Walker11, Salucci12, Breddels13} indicate that such galaxies have lower central densities and more shallow, or cored, inner density profiles. This problem of the central density of haloes is referred to as the cusp/core issue, initially identified by \cite{Flores94} and \cite{Moore94}. Specifically, at low radii, $\rm{d} \ln \rho(r) / \rm{d} \ln(r) \approx -1$ for a cuspy density profile and $\rm{d} \ln \rho(r) / \rm{d} \ln(r) \approx 0$ for a cored profile.

These discrepancies have renewed interest in the idea that dark matter may not be collisionless, but could scatter with particles of the same species. The idea of self-interacting dark matter (SIDM) was first proposed by \cite{Carlson92}, \cite{Machacek1993}, \cite{deLaix95}, and further explored by \cite{Spergel00} and \cite{Firmani00} in part to explain the cusp/core issue. In these models, the local particle scattering rate scales as
\begin{equation} \label{eq:scattering}
\Gamma(r) \propto \rho(r) \frac{\sigma}{m_x} v_{\mathrm{rms}}(r)
\end{equation}
where $\rho$ is the local density, $\frac{\sigma}{m_x}$ is the interaction cross-section per unit mass, and $v_{\mathrm{rms}}$ is the rms speed of dark matter particles. Increased scattering in high-density regions transfers energy from the outer halo to the inner halo, increasing velocity dispersion and decreasing density \citep{Burkert00, Yoshida00a, Yoshida00b, Dave01, Colin02}. Thus SIDM naturally leads to lower density cores that may better fit observations.

SIDM also provides a solution to the so-called too big to fail problem \citep{Boylan-Kolchin11, Boylan-Kolchin12}, which refers to a population of high-density subhaloes in Milky Way-like simulations with no observed analogues. If such haloes did exist as simulated, they would be too big to fail to form stars, and as such should have visible counterparts. More specifically, these haloes have a circular velocity profile which is too large to match the kinematics of known satellites. The problem exists for both the largest haloes at $z=0$, initially pointed out in \cite{Read06}, and the largest haloes at infall and before reionization. For certain cross-sections, SIDM offers a solution to this issue by reducing the circular velocity profile of all haloes \citep{Vogelsberger12, Zavala13, Vogelsberger16}.

Other solutions to the cusp/core and too big to fail issues have also been proposed that require no modifications to dark matter. Supernova driven outflows could erase cusps, as suggested and described by \cite{Navarro96,Read05, Governato10, Oh11, Governato12, Pontzen12, Teyssier13, Madau14} and \cite{Pontzen15}. In order to simultaneously solve ``too big to fail'' though, it may require an impossibly large number of supernovae \citep{Garrison-Kimmel13}. However, sufficiently early supernovae might create cores large enough to solve the problem \citep{Amorisco14}. \cite{Zolotov12} and \cite{Brooks14} also argue for early supernovae, adding in a further reduction of satellite central density due to tidal stripping to explain Milky Way observations. Additionally, late and bursty star formation could alleviate both cusp/core and too big too fail problems \citep{Onorbe15}, but not solve them entirely. Further possibilities include a combination of a reduced mass Milky Way (as proposed in \citealt{Gonzalez14}), baryon loss due to reionization and supernova outflows \citep{Sawala14}, or other combinations of these effects \citep{Pontzen14, Brook15, Wetzel16}.

In spite of several plausible purely baryonic explanations, the too big to fail problem extends to the Local Group \citep{Garrison-Kimmel14}, and may extend to the nearby field \citep{Klypin15, Papastergis15}, although that claim is sensitive to the methodology used \citep{Brook16}. Ultimately, uncertainty still remains whether or not baryonic mechanisms alone can systematically erase all concerns. \cite{Read16} for instance argue that supernovae form cores in field galaxies the size of ultrafaint dwarfs, but rely on a stellar mass to halo mass relation consistent with the abundance matching model of \cite{Behroozi13AM}, which may greatly overpredict the relationship for low-mass galaxies \citep{GarrisonKimmel14AM}. When using lower stellar masses, \cite{Penarrubia12} find no cores. Consequently, SIDM remains of interest as it could provide an attractive non-baryonic solution.

Several studies have set constraints on the strength of the SIDM interaction cross-section by searching for observable signatures. Since SIDM produces more spherical halo distributions, \cite{Miralda02} used measurements of the ellipticity of galaxy clusters to enforce a stringent constraint of $\sigma/m_x < 0.02 \, \mathrm{cm}^2/\mathrm{g}$. \cite{Randall08} set another constraint by considering the offset between the galaxy centroid and peak position of total mass in the Bullet Cluster. \cite{Yoshida00b} used simulations of core sizes in clusters, and \cite{Gnedin01} considered increased rates of subhalo evaporation due to SIDM to set constraints. A set of companion papers, \cite{Rocha13} and \cite{Peter13}, reassessed these limits and added their own constraints through studying the expected core sizes of haloes and halo shapes, respectively. They relax previous constraints, concluding that a cross-section $\sigma/m_x$ between $0.1$ and $1\, \mathrm{cm}^2/\mathrm{g}$ is sufficient to produce large enough cores in dwarfs to match observations and still withstand constraints.

At higher cross-sections, tension arises with the measured ellipticity of galaxy clusters and with overproducing cores in large haloes. These tensions can be obviated with a velocity-dependent cross-section. Motivated from particle theory, such particles could have a Yukawa potential in which the cross-section decreases for higher relative particle velocities, as proposed by \cite{Feng10, Buckley10, Loeb11}, simulated initially by \cite{Vogelsberger12}, and recently incorporated in an encompassing effective theory of structure formation by \cite{Vogelsberger16} and \cite{Racine15}. This velocity dependence results in cores on the small scales as needed, while mitigating effects on large scales where constraints are stronger \citep{Rocha13, Elbert15, Kaplinghat16}.

Since velocity-dependent cross-sections are as of yet not well constrained, and velocity-independent models are still viable, we seek to highlight a new approach where their effects could be observed. First, we investigate whether the presence of a core influences how stars in satellites of a Milky Way-like host are tidally stripped. The reduced central density of dark matter in satellites and increased central velocity dispersion both occur where stars reside. These two effects should have an impact on the stripping and distribution of stars based on the dynamics of tidal stripping and tidal steams (see \cite{Bovy14} for instance). \cite{Penarrubia10} have already demonstrated that cored satellites lose mass more quickly than cusped ones. They, however, assign cores manually and do not consider SIDM. They also emphasize tidal interactions of satellites with the disc of their host galaxy, whereas we predominantly study tidal forces arising from the host halo only.

Secondly, we investigate whether cores or subhalo evaporation change the total mass-loss rate of dark matter in satellites. Particles bound to subhaloes may be ejected after colliding with a dark matter particle in the background environment. The resulting mass-loss is called subhalo evaporation. \cite{Rocha13} demonstrate hints of this, with the subhalo maximum circular velocity function suppressed in the inner half of a host's virial radius.

Finally, we search for signatures of SIDM imprinted on the stellar halo, and observable implications of any changes in stellar and total mass-loss rates. With these goals in mind, we run simulations of four different models of SIDM to search for their impact on the disruption of satellites. By running our simulations with DM only, we isolate the effects of SIDM core formation. A complete picture would require including baryons in addition to SIDM, but such simulations are computationally more expensive and sensitive to the strength of the baryonic feedback implemented \citep{Vogelsberger14, Fry15}. As such, a suite of full hydrodynamic simulations with SIDM that resolve Milky Way-like substructure is beyond the scope of this paper.

This paper is organized as follows. Section \ref{sec:simulations} introduces the simulations, Section \ref{sec:methods} discusses our method of tagging stellar mass to dark matter particles, Section \ref{sec:data} presents the differences in stellar tidal stripping between CDM and SIDM, Section \ref{sec:subhalo_evaporation} discusses subhalo evaporation, Section \ref{sec:theory} provides a simple theory to explain and predict SIDM-driven enhancements in stellar stripping, and Section \ref{sec:implications} investigates how these in turn affect the stellar halo, stellar mass function, and half-light radii. Finally, Section \ref{sec:conclusion} summarizes our results.

\section{Simulations}
\label{sec:simulations}
We use a suite of five simulations, each of which takes its initial conditions from the Aquarius Project Aq-A-3 halo of \cite{Springel08}. Two simulations are run with a velocity-independent cross-section (the \vi models), two with a velocity-dependent cross-section (the \vd models), and one is run with pure CDM as the baseline for comparison. Using the same initial conditions allows us to isolate the effects of each SIDM model. All of the simulations are the same as those initially used in and described in \cite{Vogelsberger12}. A summary of the simulations is found in Table \ref{table:sims}. In the v-i cases, $\sigma_T^{\mathrm{max}}/m_\chi$ is the collisional cross-section per unit mass. In the v-d cases, it is the cross-section achieved when the relative particle velocity is $v_{\mathrm{max,\sigma}}$, which is the velocity at which the quantity $\sigma v$ is maximized.

\begin{figure}
\includegraphics[width=0.48\textwidth]{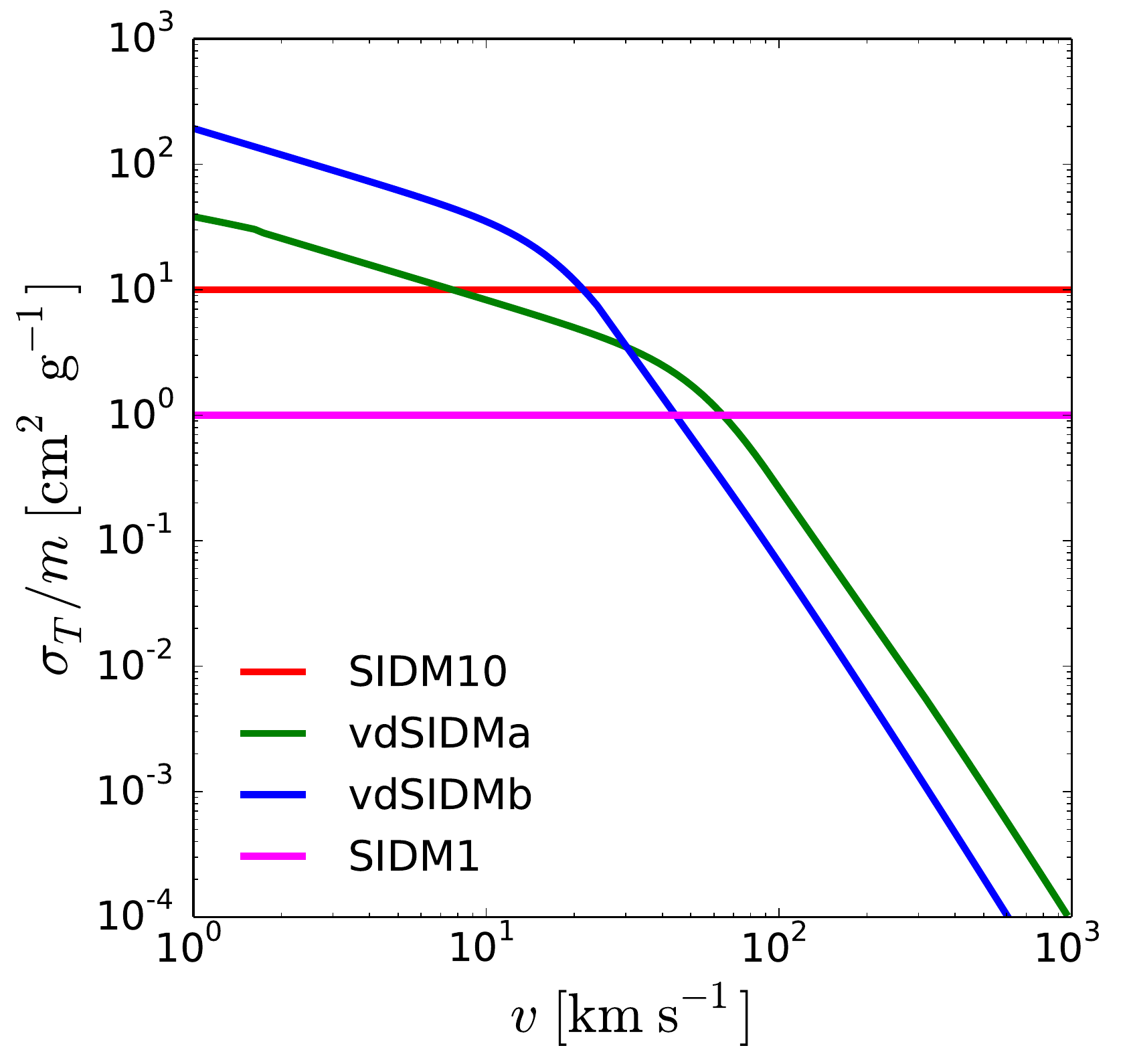}
\caption{Cross-section as a function of relative particle velocity for each SIDM model considered (see Table~\ref{table:sims}). The colour scheme used here is repeated throughout the paper to more easily identify each model.}
\label{fig:cross_sections}
\end{figure}

The transfer cross-section as a function of relative particle velocity for each of our four models is shown in Fig.~\ref{fig:cross_sections}, similar to fig. 1 of \cite{Zavala13}. Our velocity-dependent cross-sections scale approximately as $\sigma \propto v^{-4}$ at high velocity, and $\sigma \propto v^{-0.7}$ at low velocity for the range plotted. The full form of the transfer cross-section can be found in equation 1 of \cite{Vogelsberger12}. Our SIDM10 model, with a constant-cross section of $10 \, \mathrm{cm^2/g}$ is ruled out by constraints on the ellipticity of galaxy clusters, but we include it simply to more easily highlight the effects of SIDM. Each of the other three models are not ruled out, and are compatible with observed kinematics of the Milky Way dwarf spheroidal galaxies, including resolving too big to fail, as shown in \cite{Vogelsberger12}.

\begin{table}
	\tablewidth{0.38\textwidth}
	\centering	
	\caption{Summary of Simulations}
	\label{table:sims}
	\begin{tabular}{ccc} 
		\hline
		\hline
		Name &  $\sigma_T^{\mathrm{max}}/m_\chi \mathrm{(cm^2/g)}$ &  $v_{\mathrm{max,\sigma}} \mathrm{(km/s)}$\\
		\hline
		CDM &  \nodata  & \nodata \\
	    \hline
	    SIDM10 & $10$  & \nodata \\
	    \hline
 		vdSIDMa & $3.5$ & $30$ \\
		\hline
		vdSIDMb & $35$ & $10$ \\
		\hline
		SIDM1 & $1$ & \nodata \\
		\hline
		\hline
	\end{tabular}
	\tablecomments{CDM and SIDM simulations listed with cross-section parameters used in this study. $v_{\mathrm{max,\sigma}}$, is the velocity at which the quantity $\sigma v$ is maximized in the velocity-dependent cases. Naming scheme adopted from \cite{Vogelsberger13} and \cite{Zavala13} for consistency.}
\end{table}

Each simulation has a particle mass of $m_{\rm{p}} = 4.911 \times 10^{4} \, \mathrm{M_{\sun}}$, and Plummer equivalent gravitational softening length of $\epsilon = 120.5\, \mathrm{pc}$. The cosmological parameters used are $\Omega_{\rm{m}} = 0.25$, $\Omega_\Lambda = 0.75$, $h = 0 .73$, $\sigma_8 = 0.9$, and $n_{\rm{s}} = 1$.

All self-bound haloes are found using a modified version of {\sc{rockstar halo finder}} \citep{Behroozi13}. The modifications include adding full iterative unbinding to improve halo finding accuracy, as described in \cite{Griffen16}, and returning all bound particle IDs for each halo in order of boundedness. Merger trees were produced by {\sc{rockstar consistent trees}} \citep{Behroozi13MT}. We use the \cite{Bryan98} definition of the virial radius, $r_{\rm{vir}}$, which at $z=0$ for our cosmological parameters is the radius such that the mean enclosed halo density is $94.2$ times the critical density of the universe, $3H_0^2/8 \rm{\pi} G$. Unless otherwise clarified, any mention of mass refers to $m_{\rm{vir}}$, the gravitationally bound mass within $r_{\rm{vir}}$. The main halo at redshift $z=0$ in each simulation has a mass of $\sim 2.2 \times 10^{12} \, \mathrm{M_{\sun}}$, and radius of $330 \, \mathrm{kpc}$.

\section{Methods}
\label{sec:methods}
\subsection{Particle Tagging}
To study the stellar component of satellites in dark-matter-only simulations, we tag particles with stellar mass. We choose to use particle tagging over hydrodynamical simulations due to the tremendous computational time saved. This allows us to better explore SIDM parameter space via running more simulations. Additionally, while particle tagging is an imperfect method, hydrodynamical simulations are not without their own uncertainties that depend on which baryonic physics models are implemented.

In developing a tagging technique, we take guidance from a long precedent of particle tagging in simulations. Dark matter particles in the most central part of subhaloes have been assigned stellar mass at the time of subhalo accretion \citep{DeLucia08, Rashkov12}, subhalo peak mass \citep{Bailin14}, and through live tagging where mass is added at each snapshot \citep{Bullock05,Cooper10, LeBret15}. The fraction of stars tagged varies from the most bound $1\%$ \citep{Cooper10, Rashkov12, Bailin14}, to $5\%$ \citep{LeBret15}, to $10\%$ \citep{DeLucia08}. Mass may be split evenly among particles per halo \citep{DeLucia08,Rashkov12, Bailin14}, or may be split differently among dark matter particles to mimic the light profile of galaxies \citep{Bullock05, Cooper10}.

A consensus on what fraction of particles is best to tag does not exist. \cite{Bailin14} find that tagging $1\%$ versus $10\%$ results in little qualitative difference in the stellar halo, \cite{Cooper10} recommend $1-3$ per cent to avoid stellar profiles that are too concentrated or too diffuse, and \cite{LeBret15} suggest $5\%$ is better than $1\%$ to reproduce realistic stellar density profiles. Furthermore, there are cautions that particle tagging generically produces less concentrated, more structured, and more prolate galaxies \citep{Bailin14} and that live tagging schemes are superior at reproducing realistic stellar density and energy profiles \citep{LeBret15}.

While these studies demonstrate that particle tagging is an imperfect method for producing reliable quantitative results, they are still able to pick up on trends and qualitative results, which are of interest to the present study. The diversity of tagging prescriptions and conflicting recommendations on what fraction of most bound particles to tag suggests that any scheme needs to be tested within the context of the project goals.

We aim to tag enough particles for good resolution, and to probe a realistic range of where stars reside, particularly making sure to target where a core forms in SIDM. For field haloes at $z=0$, we find that cores extend to $\sim 3\%$ of the most bound particles in $10^8 \, \mathrm{M_{\sun}}$ haloes. In $10^{10} \, \mathrm{M_{\sun}}$ haloes, cores extend to $0.5\%$ in vdSIDMb on the low end, and $1.5\%$ in SIDM10 on the high end for our models (see Section \ref{sec:theory} for our calculation of core sizes). Thus, based on previous studies of tagging, to probe satellite cores, and to ensure good resolution, we would ideally tag all stars up to the $\sim 2\%$ most bound particle in all satellites, satellites of satellites, etc. throughout cosmic history. 

However, tagging particles in SIDM simulations introduces a new obstacle. Dark matter particles can elastically scatter, whereas stars should never scatter. We deal with this problem by only assigning mass to particles which do not scatter between their time of tagging and the final $z=0$ snapshot. Consequently, only a fraction of the central $2\%$ most bound particles can be tagged. In order to avoid resolution problems, we resort to still tag $2\%$ of the total bound particles, tagging all unscattered ones in order of boundedness until we reach our $2\%$. This, however, creates an additional problem. In haloes with high rates of interaction, stellar mass would be assigned at unrealistically large distances from the halo's centre. We solve this new issue by enforcing a cut-off at $5\%$. No particles that are greater than $5\%$ in their boundedness ranking can be tagged. If fewer than $2\%$ of the total number of halo particles can be tagged within the interval of the $5\%$ most bound particles, then fewer than $2\%$ are tagged. This represents a compromise between resolution and tagging the truly central regions of satellites. 

\begin{figure*}
  \centering
  \begin{tabular}[b]{@{}p{0.55\textwidth}@{}}
    \centering\includegraphics[width=.55\textwidth]{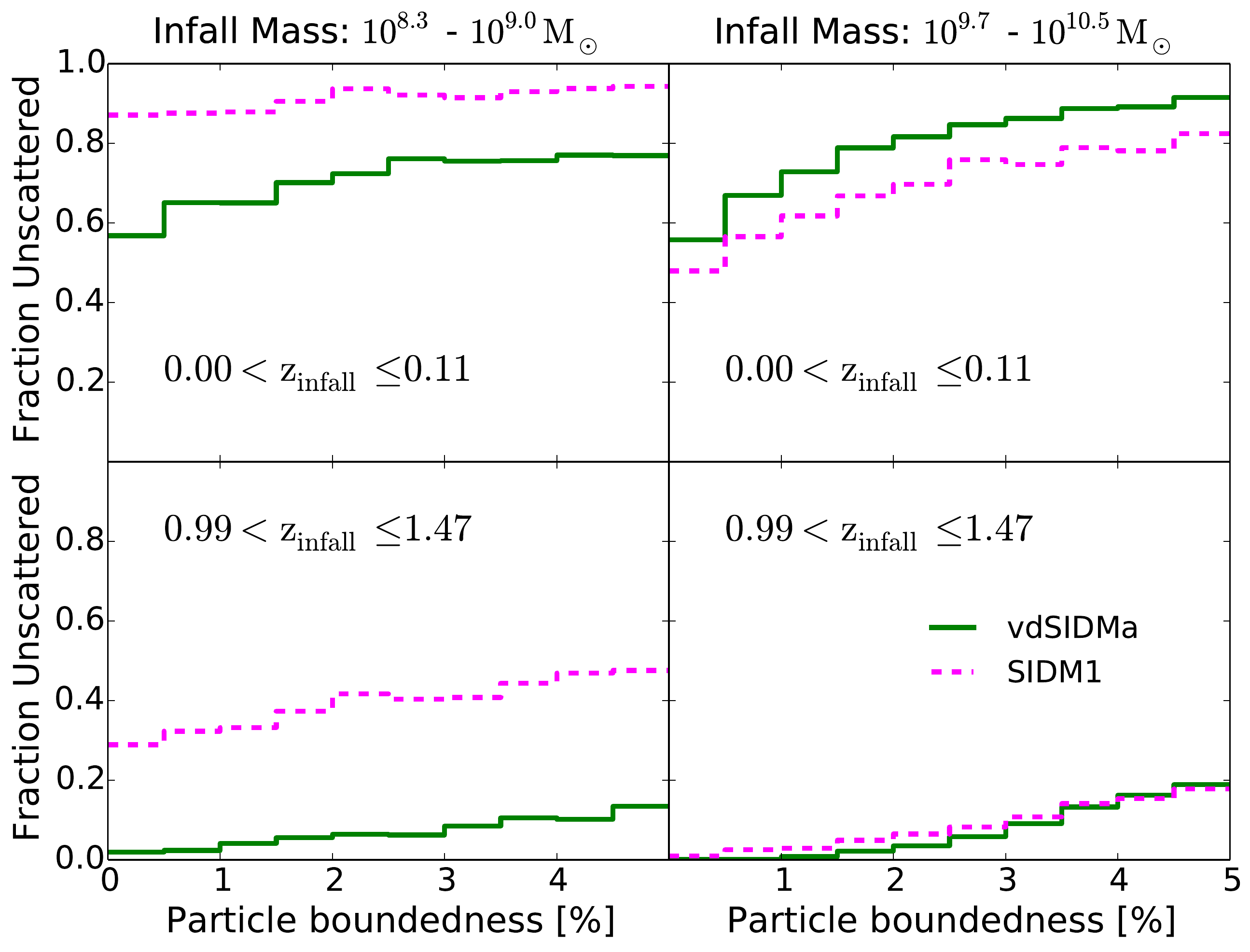} \\
    \centering\small
  \end{tabular}%
  \quad
  \begin{tabular}[t]{@{}p{0.40\textwidth}@{}}
    \centering\includegraphics[width=.41\textwidth]{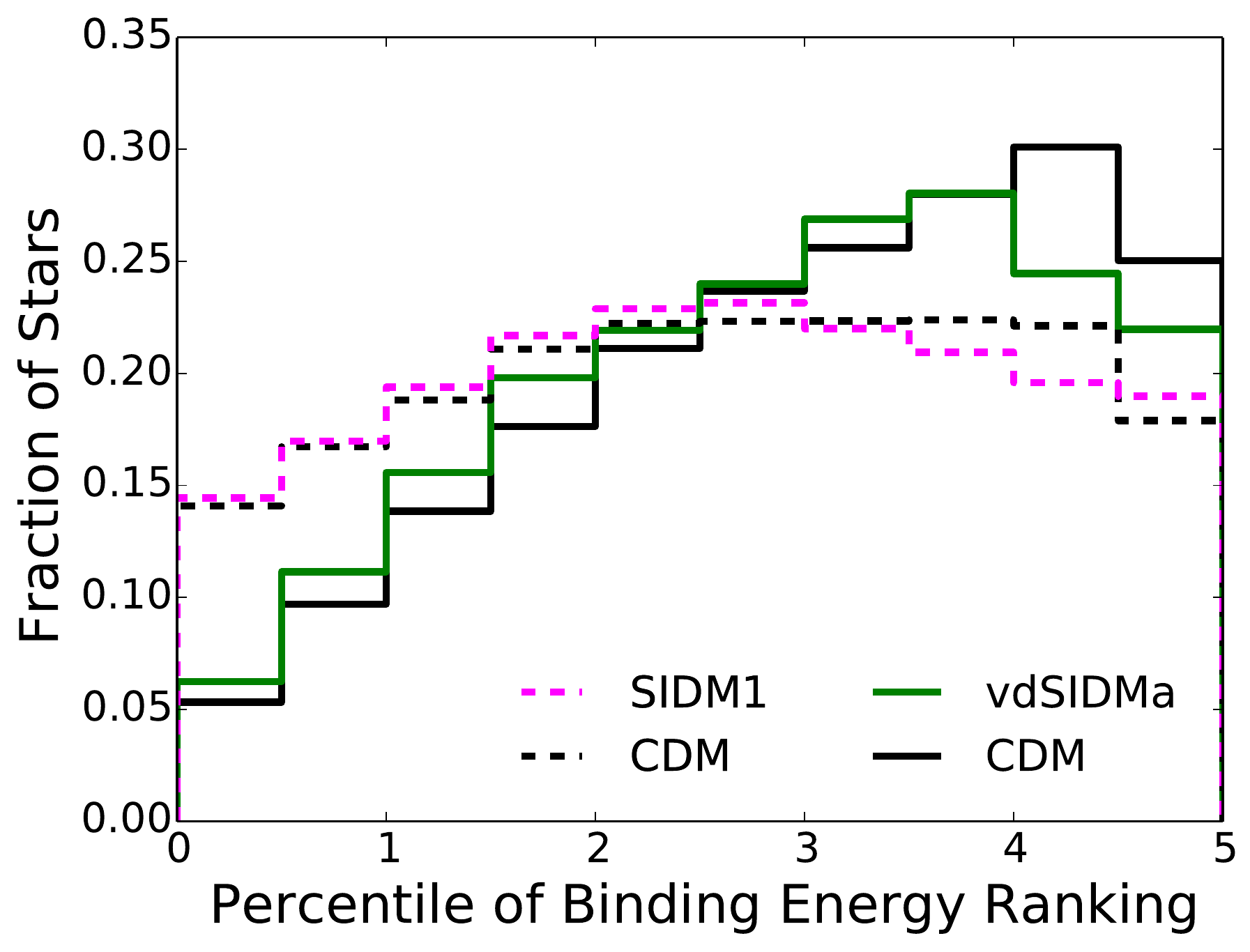} \\

    \centering\small 
  \end{tabular}
  \caption{Left-hand panel: distribution of the fraction of particles that do not scatter between infall and $z=0$ (and can thus be tagged) as a function of their rank order boundedness in satellites at infall. A sample of two infall time bins and two infall mass bins are shown for the SIDM1 simulation in magenta (dashed lines), and the vdSIDMa simulation in green (solid lines). See Table~\ref{table:sims} for differences between the SIDM models. The unscattered distributions of all satellites within an infall mass and time bin are averaged to produce a look-up table used to tag the corresponding CDM case. Right-hand panel: the normalized distribution of the fraction of tagged particles per boundedness percentile at the time of tagging for all tagged particles in the simulation. We verify that our tagging of the CDM simulations mirror that of the target SIDM simulation. The dashed-line CDM distribution mirrors that of the corresponding SIDM1 distribution (dashed magenta line) and the solid-line CDM distribution mirrors that of the corresponding vdSIDMa distribution (solid green line).}
  \label{fig:tag_distr_grid}
\end{figure*}

The distribution of particles that remain unscattered within the $5\%$ most bound interval varies between each SIDM simulation, and as a function of halo mass and infall time. For an unbiased comparison with CDM, it is imperative to tag particles in CDM in as similar a fashion as possible to the corresponding SIDM case. We divide all haloes into bins according to their infall time and infall mass. Each halo has a distribution recorded for it, cataloguing the fraction of particles that remain unscattered as a function of rank order particle boundedness. We consider particles, starting with the most bound, in intervals of $0.5\%$ of the total number of bound particles, recording the fraction of unscattered particles in each interval. The distributions of all haloes within an infall scale and mass bin are then averaged. An example of four averaged distributions for the SIDM1 and vdSIDMa simulations are shown in the left-hand panel of Fig.~\ref{fig:tag_distr_grid}. As expected, satellites which are more massive (higher central density) and fall in earlier experience more scattering and have fewer remaining particles to tag. This procedure results in a look-up table used to tag the CDM simulation in four ways, once for each of the four SIDM simulations. An infalling satellite in the CDM simulation is tagged to reproduce the right look-up table distribution, with specific particles within a particle boundedness interval being selected uniformly at random. In cases such as the $0.99 <\mathrm{z_{infall}} < 1.47$, $10^{9.7} \, \mathrm{M_{\sun}} < M < 10^{10.5} \, \mathrm{M_{\sun}}$ bin in the left-hand panel of Fig.~\ref{fig:tag_distr_grid}, fewer than $2\%$ of all bound particles can be tagged on average. Correspondingly, such haloes in the CDM simulation are tagged with fewer than $2\%$ of all particles. 

After completing this procedure, we verify that our tagging did not introduce significant biases. One particular test is shown in the right-hand panel of Fig.~\ref{fig:tag_distr_grid}. This shows the normalized distribution of the boundedness percentile at which particles were tagged for all tagged particles in all subhaloes. We present the cases of vdSIDMa (solid green line), SIDM1 (dashed magenta line), and each of their corresponding CDM instances (solid black and dashed black lines respectively). Whereas the left-hand panels shows the unnormalized distribution of unscattered particles, this panel shows the normalized distribution of actually tagged particles. Importantly, the CDM simulation is tagged in different ways to mimic the features of the different SIDM cases. The dashed CDM line mirrors that of SIDM1, and the solid CDM line mirrors that of vdSIDMa. Imperfections in the match are due mostly to slight differences in merger history. 

Other tagging attempts which did not produce matches between CDM and SIDM pairs resulted in biases, such as stars tagged at systematically larger radii in SIDM than in CDM, resulting in them being more easily tidally stripped. For full confidence in our tagging method, we perform a variety of additional tests verifying that our primary results are not simply an artefact of our tagging technique. A summary of these robustness tests can be found in Appendix~\ref{sec:tagging_tests}.

Upon selecting particles for tagging, we assign to them each a stellar mass according to the abundance matching prescription of stellar mass to halo mass in \cite{Moster13}. Since this defines a relationship for haloes at infall, we tag particles at infall. Infall is the snapshot before a halo first becomes a subhalo of the main host. A subhalo is defined as a halo that resides within the virial radius of its larger host. Given the virial mass of a halo at infall, we use the abundance matching relationship to specify total stellar mass, then split that evenly among particles tagged within the halo. This means particles within a halo have the same stellar mass, but particles in different haloes have different stellar masses.

At a halo mass of $10^8 \, \mathrm{M_{\sun}}$, the stellar mass is only $80 \, \mathrm{M_{\sun}}$. Due to this negligible contribution, we do not tag haloes with an infall mass below that value. In fact, the more recent stellar mass to halo mass investigation by \cite{Brook2014} suggests that $80 \, \mathrm{M_{\sun}}$ is a great overestimate. They find a much steeper slope going towards lower stellar masses below a halo mass of $10^{10.3} \, \mathrm{M_{\sun}}$. Models by \cite{Behroozi13} and \cite{Sawala15} on the other hand, predict much higher stellar mass to halo mass values. \cite{Sawala15} for instance predicts a stellar mass of $8 \times 10^4 \, \mathrm{M_{\sun}}$ for $10^8 \, \mathrm{M_{\sun}}$ DM haloes at $z=0$ that host galaxies, but argue that nearly all haloes that size and smaller remain dark since reionization suppresses star formation entirely. None the less, the uncertainty in what stellar mass to assign to low mass galaxies does not affect our primary results in Sections \ref{sec:data}-\ref{sec:theory} since they are based on the ratio of mass in a satellite after infall to the mass at infall, which is independent of the stellar mass to halo mass ratio used. Uncertainty in this relationship does somewhat impact the implications of our findings in Section~\ref{sec:implications}, and is accordingly discussed there.

\begin{figure*}
  \hspace{-15mm}
  \begin{tabular}[b]{@{}p{0.20\textwidth}@{}}
	\includegraphics[width=.25\textwidth]{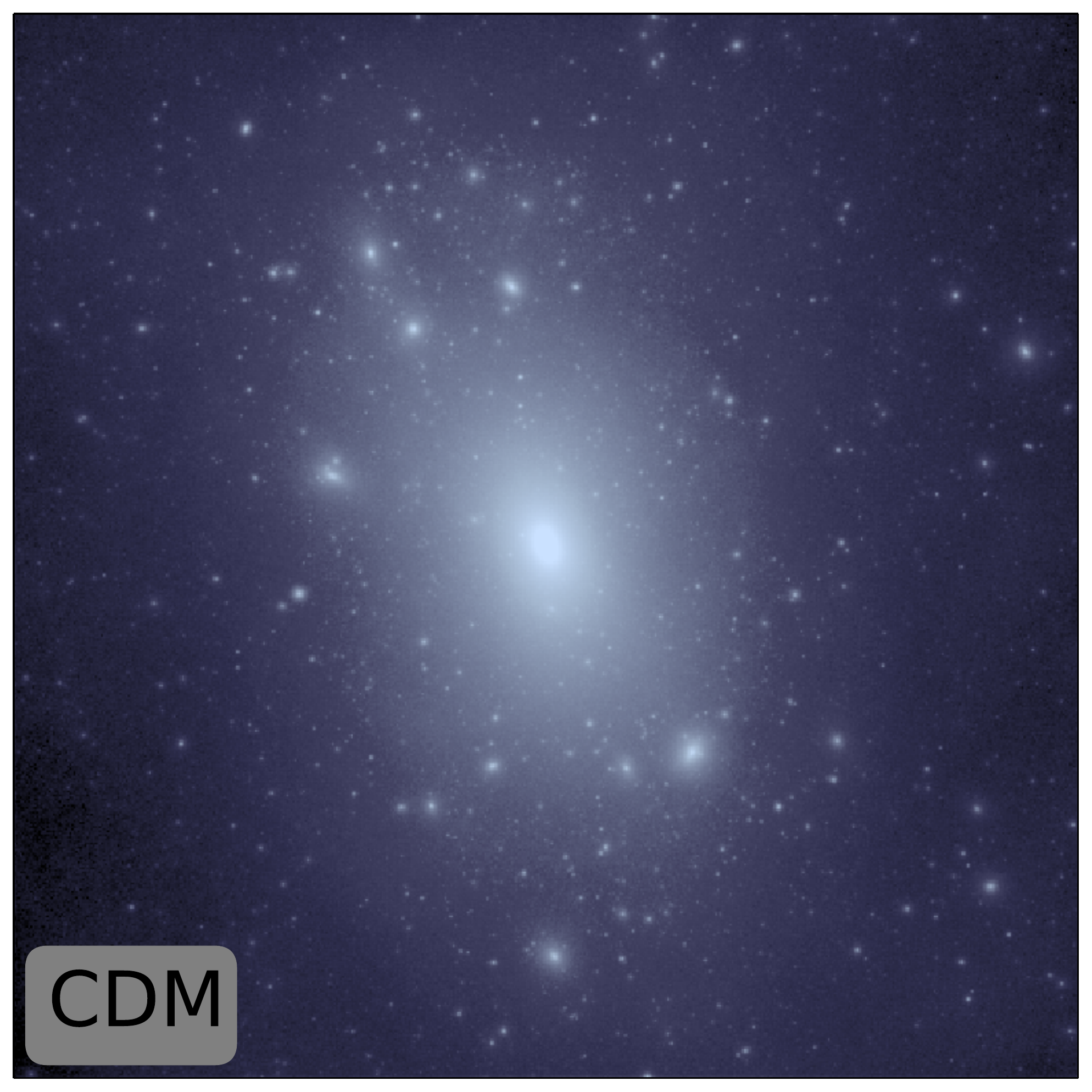} \\
  \end{tabular}%
  \quad
  \begin{tabular}[t]{@{}p{0.20\textwidth}@{}}
  \includegraphics[width=.25\textwidth]{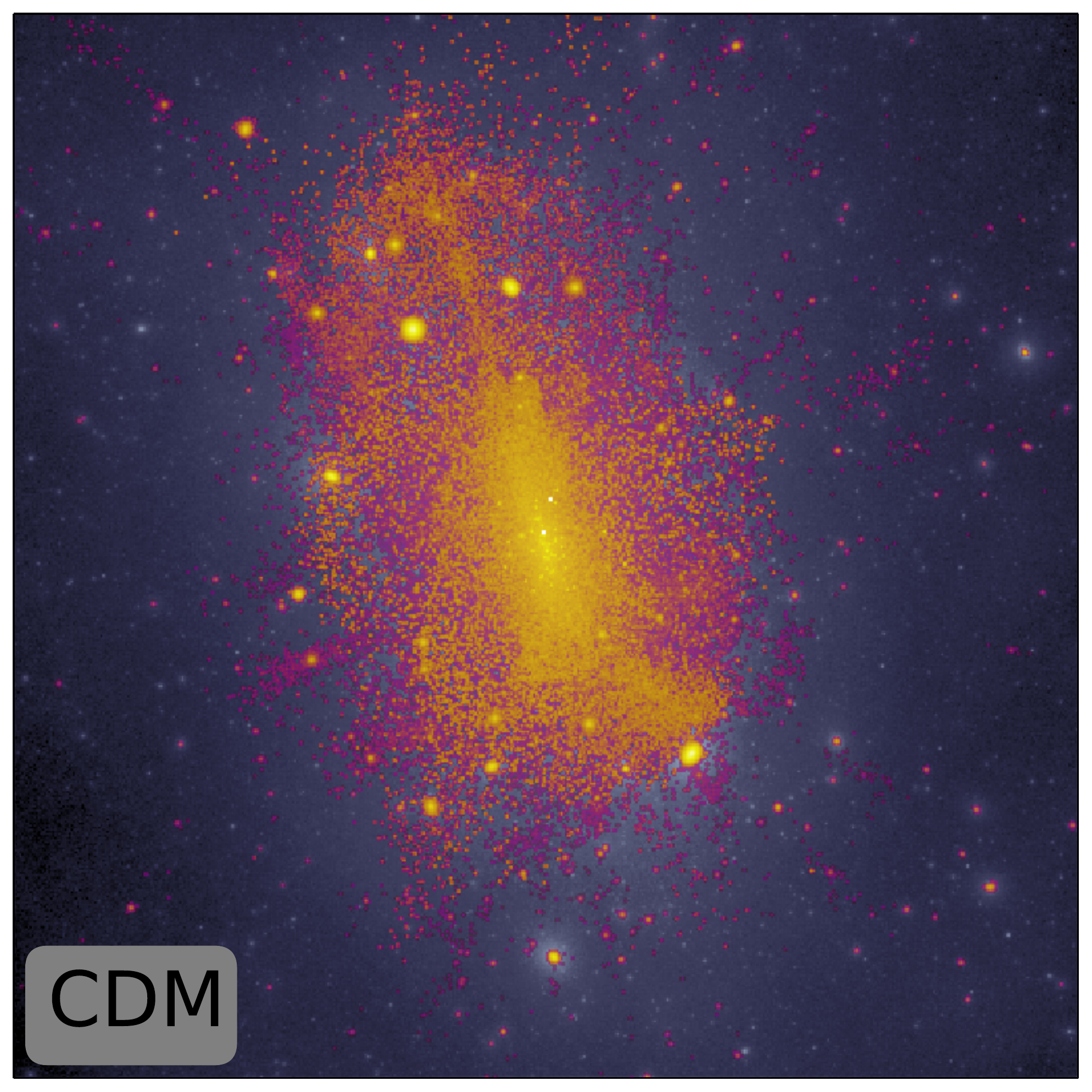} \\
  \end{tabular}
    \quad
  \begin{tabular}[t]{@{}p{0.20\textwidth}@{}}
   \includegraphics[width=.25\textwidth]{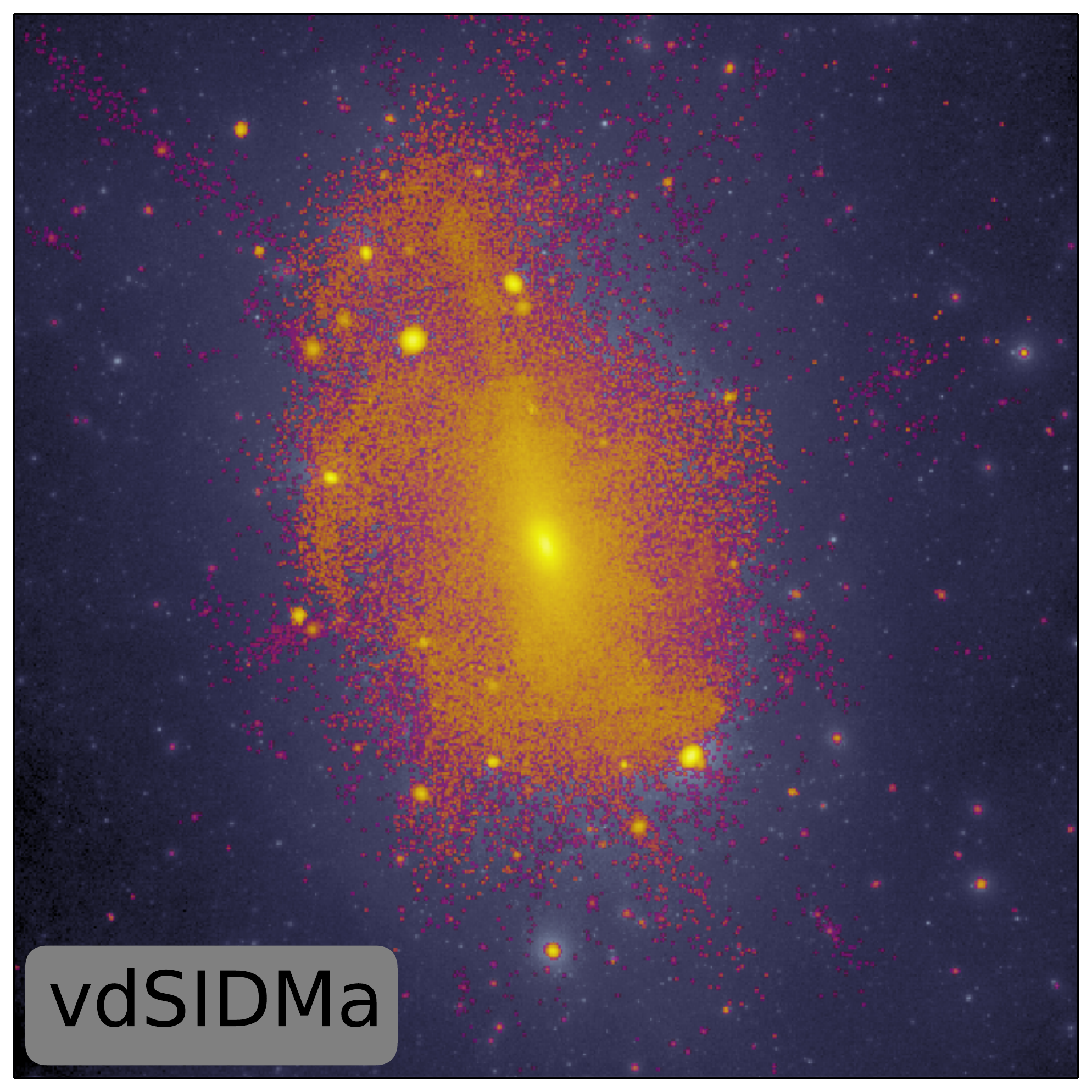} \\
  \end{tabular}
    \quad
  \begin{tabular}[t]{@{}p{0.20\textwidth}@{}}
   \includegraphics[width=.25\textwidth]{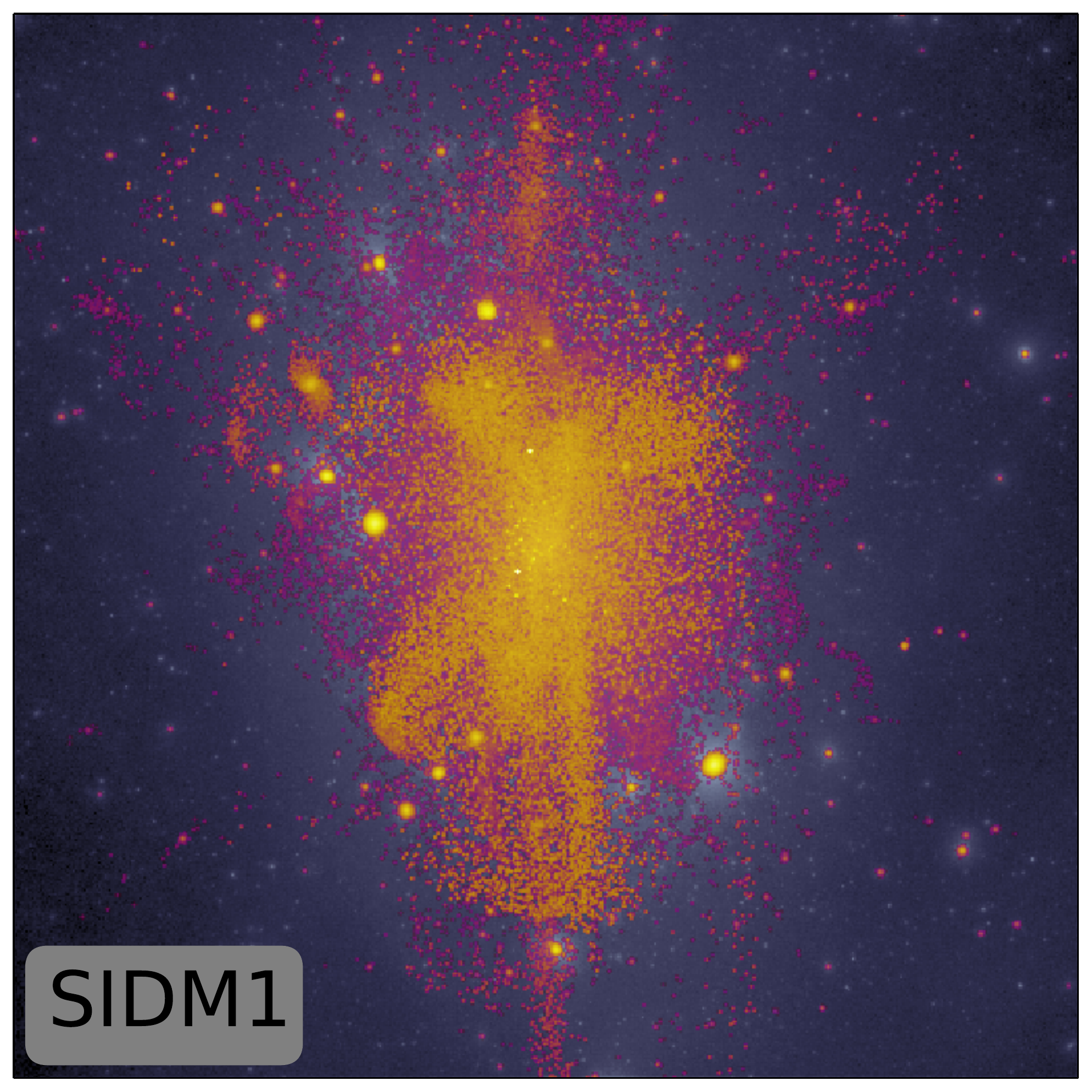} \\
  \end{tabular}
  \caption{Logarithmic density projections of dark matter and tagged stellar mass in our simulations. From left to right: dark matter in the CDM simulation, dark matter with overlaid tagged stellar mass in the CDM simulation, dark matter with overlaid tagged stellar mass in the vdSIDMa simulation, dark matter with overlaid tagged stellar mass in the SIDM1 simulation. Due to core formation in the host of SIDM1, orbits are altered relative to the CDM simulation and the stellar halo deviates in appearance from that of CDM. In vdSIDMa, only a small core forms in the host, resulting in little difference in the stellar halo's appearance. All images have a box length of $500 \, \mathrm{kpc}$.}
  \label{fig:density_projections}
\end{figure*}

A visual of a sample of our haloes with and without stellar tagging is shown in Fig.~\ref{fig:density_projections}. Dark matter is shown from low to high density in blue to silver, and stars in magenta to yellow.

\section{Stellar Stripping}
\label{sec:data}
\begin{figure*}
\centering
\includegraphics[width=0.68\textwidth]{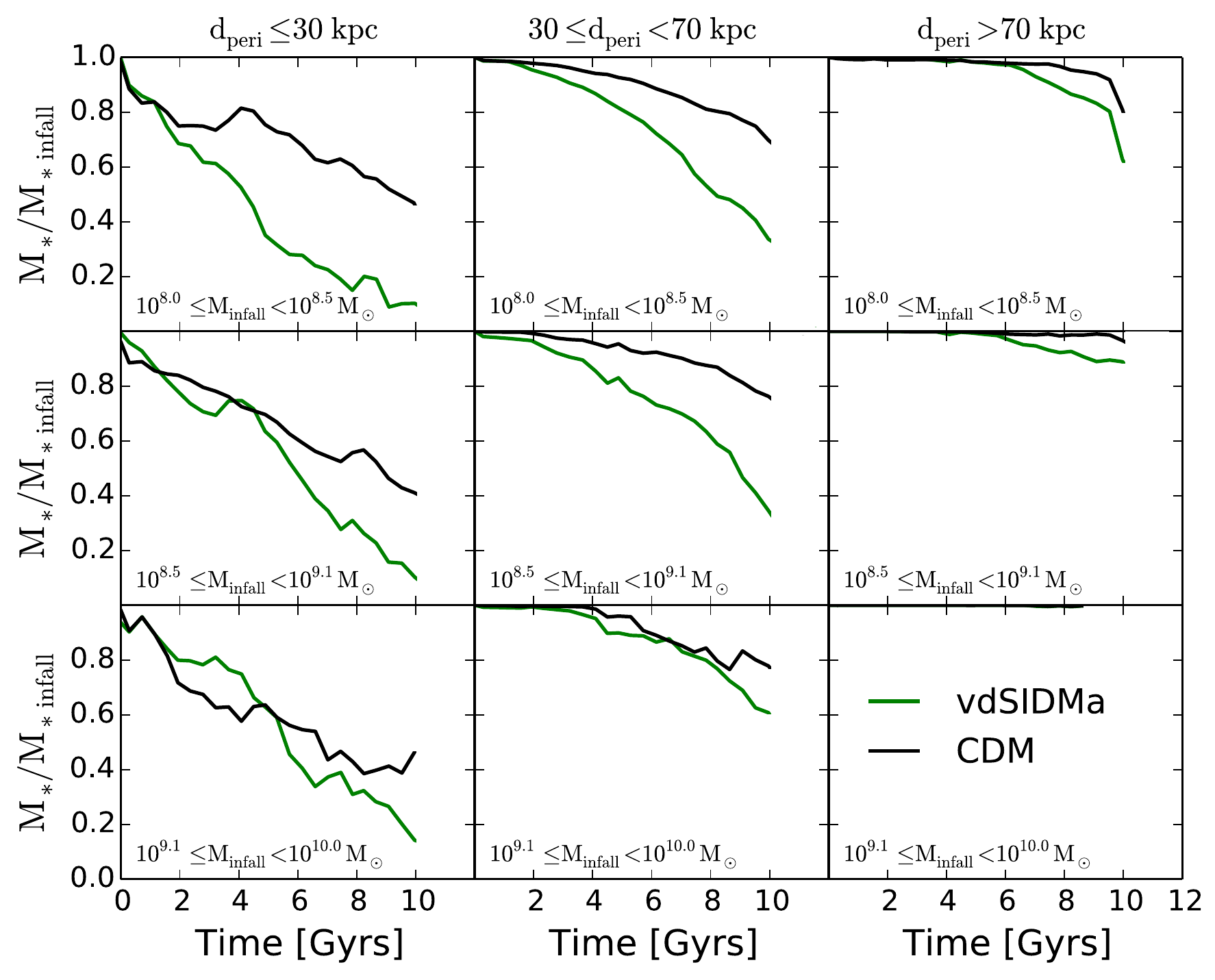}
\caption{Fraction of stellar mass remaining in satellites as a function of time since infall, averaged over all satellites for vdSIDMa compared to CDM. Satellites are divided into different bins according to their mean pericentre distance, $d_{\rm{peri}}$ (columns), and infall mass, $M_{\rm{infall}}$ (rows). Satellites with $d_{\rm{peri}} < 30 \, \mathrm{kpc}$ would pass through the host's disc in a real Milky Way analogue, but do not in our dark-matter-only simulations. Stripping in both CDM and SIDM should therefore be increased in this interval. Stellar mass stripping is enhanced in the vdSIDMa model relative to CDM for all orbit bins, particularly for lower-mass satellites.}
\label{fig:strip_v_time_p2}
\end{figure*}

\begin{figure*}
\centering
\includegraphics[width=0.68\textwidth]{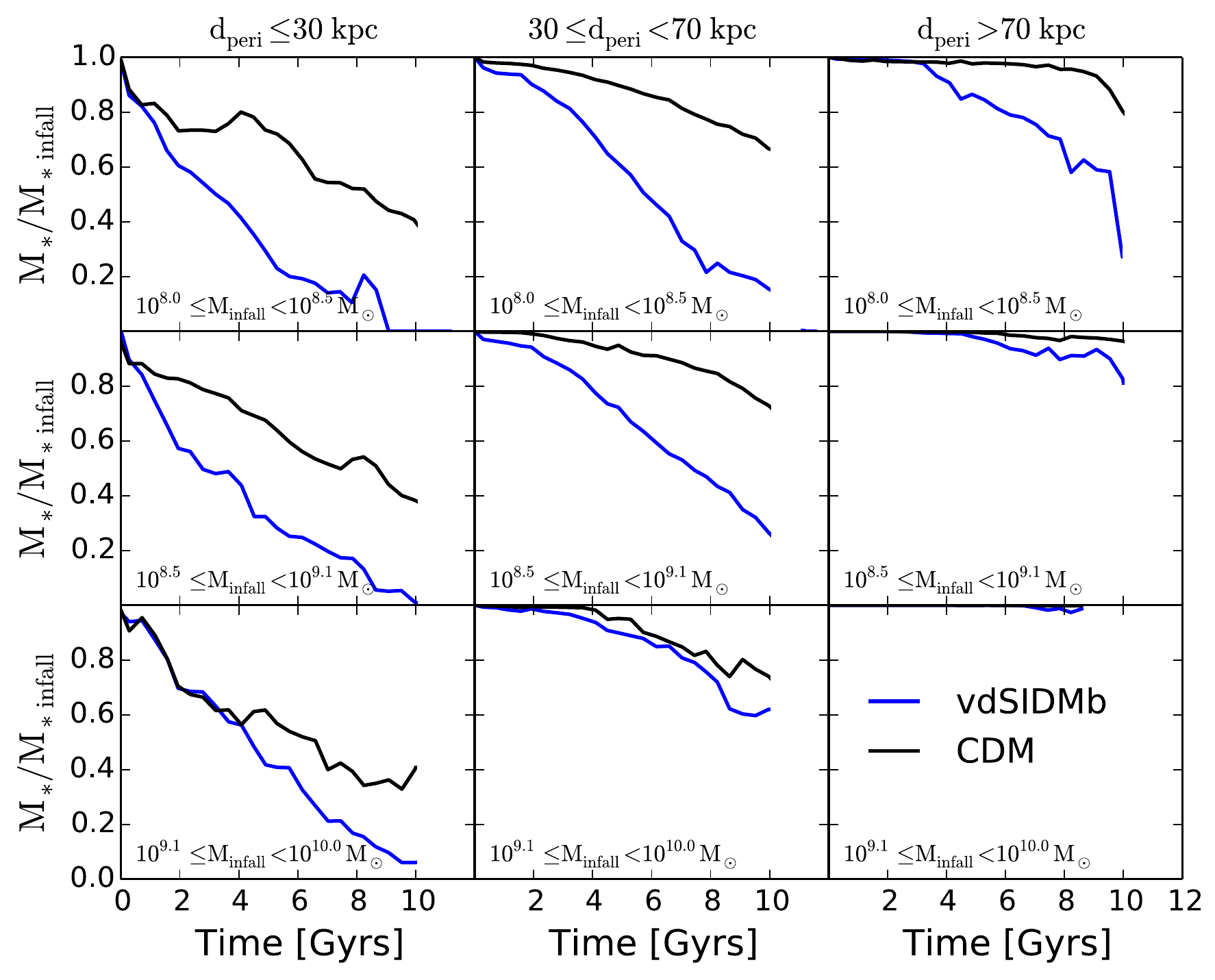}
\caption{Same as Fig.~\ref{fig:strip_v_time_p2}, except for model vdSIDMb. Stellar mass stripping is enhanced for all orbit bins in vdSIDMb relative to CDM, with a strong dependence on satellite mass. The effect is the largest of all our models for low-mass haloes, and the smallest for high-mass haloes.}
\label{fig:strip_v_time_p3}
\end{figure*}

\begin{figure*}
\centering
\includegraphics[width=0.68\textwidth]{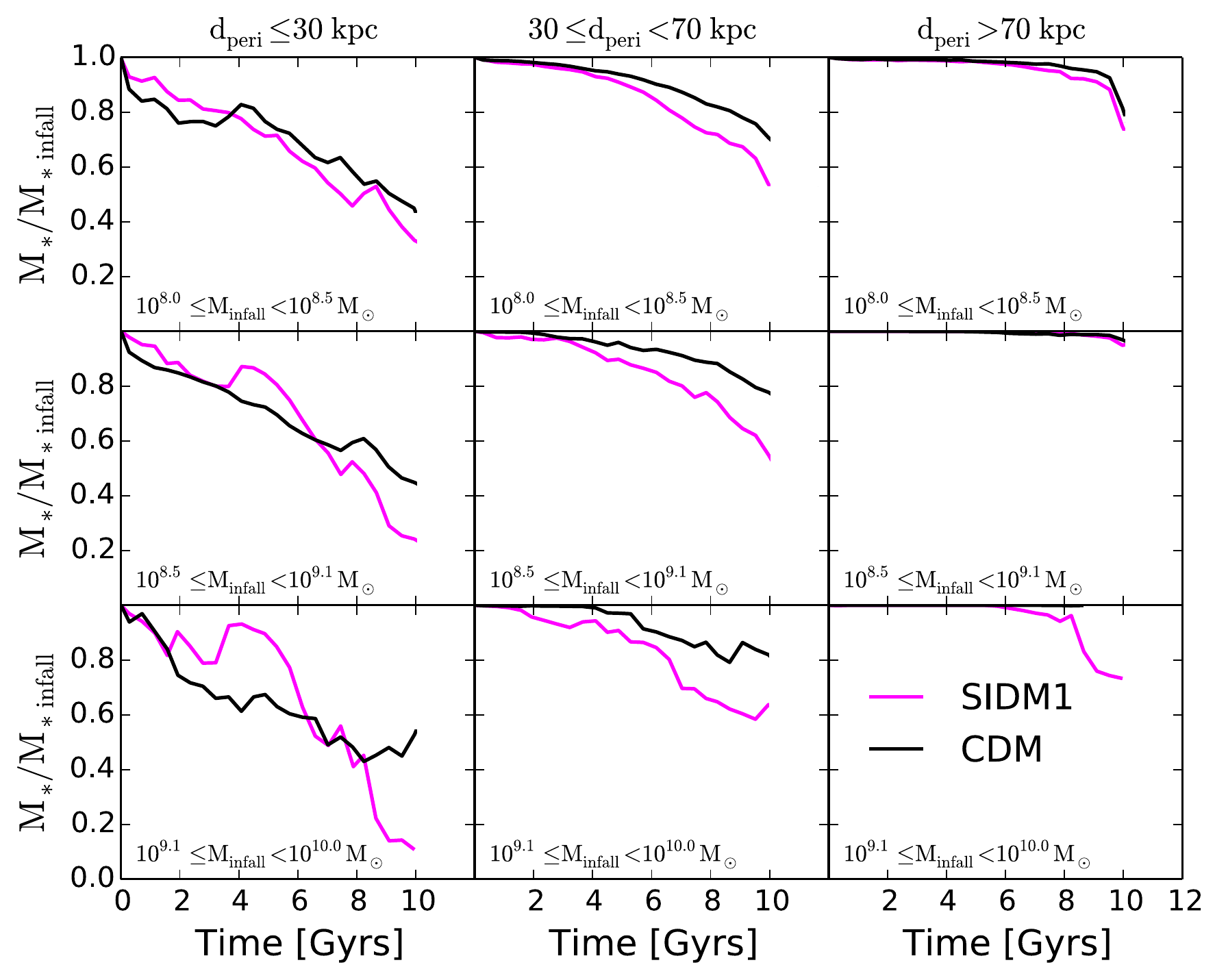}
\caption{Same as Fig.~\ref{fig:strip_v_time_p2}, except for model SIDM1. Stellar mass stripping is enhanced in SIDM1 relative to CDM for orbit bins with $d_{\rm{peri}} > 30 \, \mathrm{kpc}$, with a weak dependence on satellite mass. More massive satellites experience a greater enhancement. Below $30 \, \mathrm{kpc}$, a reduced tidal field due to the host halo's core in SIDM counteracts the effects of a core in the subhaloes. The combined effect makes stripping looks similar to CDM.}
\label{fig:strip_v_time_p4}
\end{figure*}

Using the particle tagging technique described in Section \ref{sec:methods}, we quantify the stripping of stars in each of our SIDM models compared to the baseline CDM simulation. In Figs \ref{fig:strip_v_time_p2}-\ref{fig:strip_v_time_p4}, we present the fraction of stellar mass remaining in satellites as a function of time since infall for the models vdSIDMa, vdSIDMb, and SIDM1, respectively. Since stripping will depend on the satellite orbit, infall time, and mass, we divide our samples into bins of different infall mass and mean pericentre achieved. The fraction of stars remaining versus time is then averaged over all subhaloes within each bin.

All satellites that fell into the host, whether extant at $z=0$ or destroyed, are included in the sample. In total, there are around $800$ satellites considered per simulation. Most satellites eventually merge with the host, i.e. they are no longer found by the halo finder, or fall in more recently than $10$ Gyr ago. Consequently, as time increases, there are fewer and fewer remaining satellites to average over for each data bin. We enforce a cut-off beyond $10$ Gyr where the data become particularly limited.

The mean pericentre of satellite orbits is well correlated with the strength of disruptive tidal forces exerted on a satellite, even though it ignores the complexity of a full orbit. We therefore divide our sample into bins of $0-30 \, \mathrm{kpc}$, $30-70 \, \mathrm{kpc}$, and $> 70 \, \mathrm{kpc}$ mean pericentre crossing. We choose $30 \, \rm{kpc}$ as the upper limit of the first bin since it corresponds approximately to the radius at which a Milky Way-like disc causes divergent tidal stripping in cored versus cuspy subhaloes as studied in \cite{Penarrubia10}. Both cored and cuspy satellites experience more disruption with a disc, but cored satellites with low pericentres experience a much greater increase in disruption than cuspy ones. Since our simulations do not include a disc in the host, the effects of stripping in both CDM and SIDM cases in the $d_{\rm{peri}} < 30\, \mathrm{kpc}$ bin will be underestimated relative to the other bins.

We choose our infall mass bins as $10^{8.0} \, \mathrm{M_{\sun}} < M_{\rm{infall}} < 10^{8.5} \, \mathrm{M_{\sun}}$, $10^{8.5} \, \mathrm{M_{\sun}} < M_{\rm{infall}} < 10^{9.1} \, \mathrm{M_{\sun}}$, and $10^{9.1} \, \mathrm{M_{\sun}} < M_{\rm{infall}} < 10^{10} \, \mathrm{M_{\sun}}$ to highlight interesting ranges of the halo mass dependence for each SIDM model, and to have enough satellites per bin to see trends of the average effect. Specifically, the mass $M_{\rm{infall}} < 10^{9.1} \, \mathrm{M_{\sun}}$ is chosen to highlight a transition in tidal effects on the v-d models from significant to insignificant. On halo mass scales above $10^{10} \, \mathrm{M_{{\sun}}}$, there are only $\sim 10$ subhaloes that ever enter the host. In these cases, the specific orbits of the satellites drive the  differences in evolution more than SIDM, and cannot be averaged out. Additionally, dynamical friction drags massive haloes more rapidly to low pericentre orbits, leaving few examples with large orbits.

Due to the difficulty of particle tagging in a simulation with scattering particles, there are a few considerations that must be taken into account.

First, in the model SIDM10, there are too few non-scattered particles to tag as stars to extract meaningful results. Our goal of tagging $2\%$ of all bound particles within the interval of the $5\%$ most bound particles can only be achieved for satellites falling into the host within $\sim 0.9$ Gyr of $z=0$. Tagging $1\%$ of particles is no longer possible for infalls earlier than $\sim 2.5$ Gyr before $z=0$. At earlier times, the fraction of particles that can be tagged quickly approaches $0\%$. We therefore do not include SIDM10 in our results here, but expect that its effects will be a more extreme version of SIDM1. 

Secondly, due to the different distributions of particles tagged in each SIDM model, it is important to only make comparisons with the CDM data tagged in accordance with the corresponding SIDM model. In spite of this caution, the stripping of particles in CDM is not substantially altered between different tagging instances, as seen by the black lines in Figs \ref{fig:strip_v_time_p2}-\ref{fig:strip_v_time_p4}. Any differences that do exist are small compared to the difference in stripping due to the SIDM effects. This indicates that primarily dark matter physics drives the signals we study, not our tagging technique.

Thirdly, we combine all satellites regardless of infall time. Satellites with an earlier infall time have systematically higher rates of stripping in both SIDM and CDM. This is not physical, but instead due to particles being tagged further from the subhalo centre as compared to recently accreted satellites. Fewer unscattered particles remain in satellites that fall in early since there is more time for scattering to occur between infall and $z=0$. Consequently, we must go to larger distances on average to find unscattered particles. In spite of this effect, the relative difference between SIDM and CDM stripping remains the same independent of infall time. It does, however, mean that the fraction of stars stripped in our simulation data is likely an overestimate of the fraction that would be stripped in a real Milky Way satellite.

With concerns of tagging artefacts assuaged, we look at the physical effects of SIDM. In the v-d models, we see a large effect of SIDM enhancing tidal stripping relative to CDM for low mass haloes, and the effect diminishing for higher mass haloes. The high cross-section in vdSIDMb at low relative velocities typical for haloes with $M_{\rm{infall}} \le 10^{8.5} \, \mathrm{M_{\sun}}$ (for $v_{\rm{max}} \le 13.5 \, \mathrm{km/s}$, $\sigma/m_x \ge 25.6 \, \mathrm{cm^2/g}$), results in a very large effect, even for haloes with a mean pericentre $> 70 \, \mathrm{kpc}$, where tidal forces are small. The strength of stripping for halo masses below $10^{9.1} \, \mathrm{M_{\sun}}$ in vdSIDMa is weaker in comparison, particularly for mean pericentres above $30 \, \mathrm{kpc}$. 

In SIDM1, for our intermediate range of mean pericentres, there is a clear enhancement of tidal stripping relative to CDM. This effect gradually increases with larger mass haloes. Furthermore, this enhancement overtakes the effect seen in vdSIDMa and vdSIDMb for haloes between $10^{9.1} \, \mathrm{M_{\sun}}$ and $10^{10} \, \mathrm{M_{\sun}}$.

For haloes in SIDM1 with mean pericentre above $70 \, \mathrm{kpc}$, there is neither appreciable stripping in SIDM nor in CDM. For haloes with mean pericentre below $30 \, \mathrm{kpc}$, we see an interesting effect where self-interactions do not cause any additional enhancement in stripping. This is dramatically different than the velocity-dependent models where a strong enhancement is apparent. A large core forms in the host in SIDM1, reducing the local tidal field that causes stripping, whereas only a small core forms in the host of the v-d models.

Extending the trend of diminishing enhancement of stripping towards higher halo masses in the v-d models, there should be little difference between SIDM and CDM for haloes above $10^{10} \, \mathrm{M_{{\sun}}}$. Indeed, by matching individual haloes to each other via identifying common particle IDs between simulations, we find no significant difference in stripping in the few haloes that follow similar orbits. For SIDM1, the opposite effect of enhanced stripping increasing towards higher masses exists. We therefore expect some difference to exist for haloes above $10^{10} \, \mathrm{M_{{\sun}}}$, but the SIDM effect is sub-dominant to the more critical orbital parameters and infall time. By matching large subhaloes to each other in SIDM1 and CDM, we could not isolate a clear SIDM related effect. The implications of stellar stripping in these largest systems on the stellar halo are discussed in Section \ref{sec:stellar_halo}.

\section{Subhalo Evaporation and DM stripping}
\label{sec:subhalo_evaporation}
Historical interest in SIDM stems from subhalo evaporation. As a satellite orbits within the background dark matter in the host, high velocity collisions with background particles can eject subhalo particles. Consequently, subhalo evaporation could suppress the subhalo mass function and alleviate the missing satellite problem, as originally proposed in \cite{Spergel00}. 

We therefore investigate the strength of subhalo evaporation in our models by studying the mass-loss rate of the total bound mass of subhaloes. In turn, we assess whether it plays an important role in causing the increased rate of stellar mass-loss in SIDM already reported in Section \ref{sec:data}. Significant subhalo evaporation can reduce the total mass of a satellite, lowering its binding energy, and increasing the likelihood of stars being stripped. However, tidal forces alone could also increase the total mass-loss rate of cored SIDM satellites relative to cuspy CDM ones, and we must disentangle the two effects. Fortunately, as will become clear, we see no significant effect of tidal forces causing divergent total mass-loss rates for the subhaloes we study, and can therefore isolate the effects of subhalo evaporation.

Previously, \cite{Vogelsberger12} studied the effect of subhalo evaporation on the subhalo mass function in the models SIDM10, vdSIDMa, and vdSIDMb and found no change in the two v-d models. They did, however, find mass function suppression in SIDM10, especially for subhaloes closer to the host's centre. \cite{Rocha13} also investigated subhalo evaporation, but in the context of the cumulative maximum circular velocity function. They find a small suppression in SIDM1 for subhaloes in $10^{12} \, \mathrm{M_{\sun}}$ hosts, again magnified for subhaloes in the inner regions.

\begin{figure*}
\centering
\includegraphics[width=0.68\textwidth]{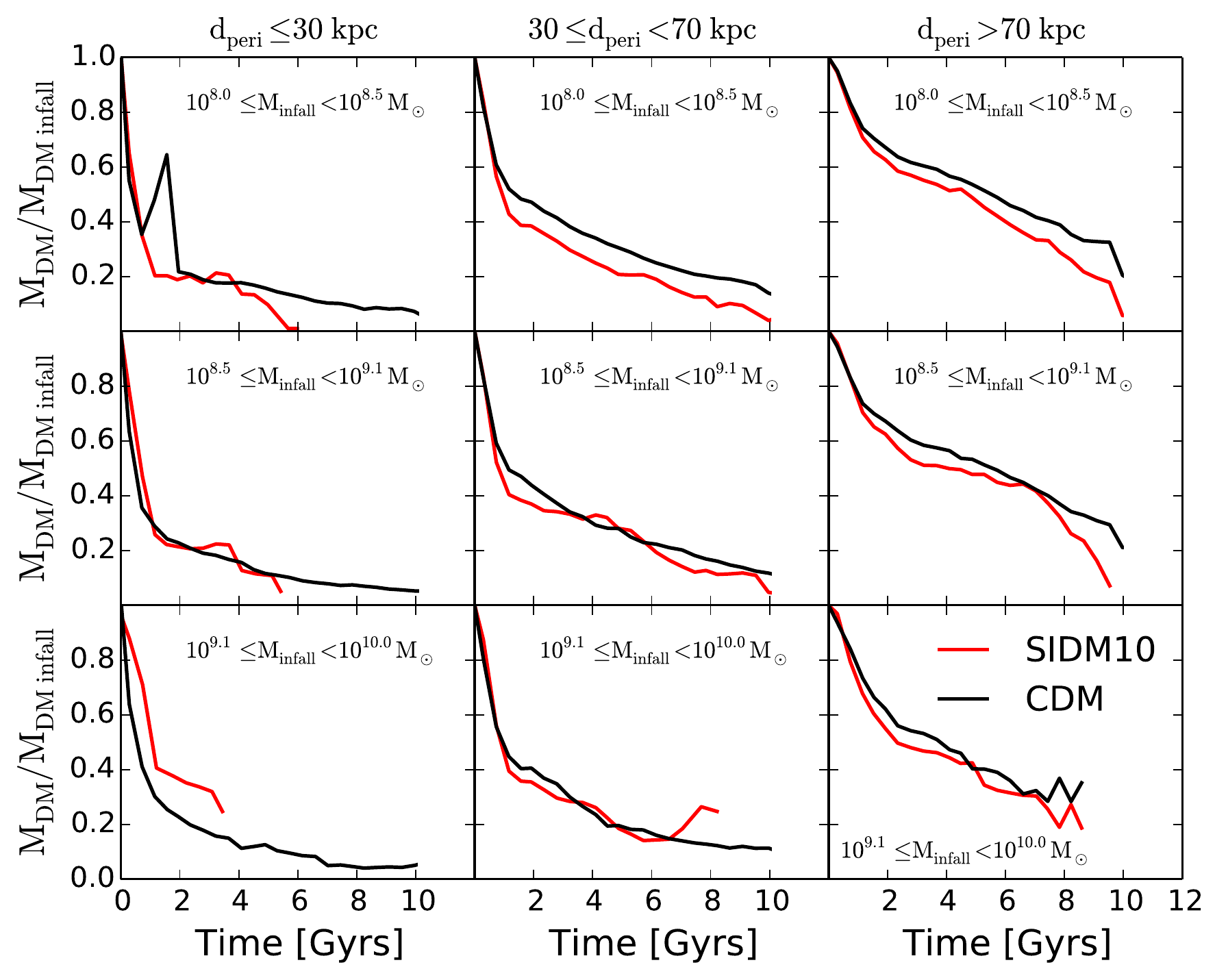}
\caption{Total mass remaining in satellites over time binned by mean pericentre, $d_{\rm{peri}}$, and infall mass, $M_{\rm{infall}}$. Scattering is high enough for subhalo evaporation to reduce the mass of satellites in SIDM10 faster than in the CDM case.}
\label{fig:allmass_strip_v_time_p1}
\end{figure*}

We expand upon these studies, looking at total mass-loss over time as a function of mean orbital pericentre, $d_{\rm{peri}}$, and mass, $M_{\rm{infall}}$, as in Section \ref{sec:data}, instead of a static $z=0$ subhalo count. Since we track all mass, not just stellar mass, particle tagging does not affect the results. Whereas previously we had to eliminate haloes in which no unscattered particles remained for tagging, now we can include all haloes in our sample. Consequently, we can study the SIDM10 model, shown in Fig.~\ref{fig:allmass_strip_v_time_p1}. Here, we see significant effects of subhalo evaporation. In all cases where $d_{\rm{peri}} > 30$ kpc, there is a clear and consistent break from CDM $1 \, \mathrm{Gyr}$ after infall. In the lowest mass bins, the separation continues with SIDM haloes having an additional $10\%$ of their original mass lost. By $10 \, \mathrm{Gyr}$, this translates into having only half of the mass of their CDM counterparts. In the highest mass bins, there are hints of increased mass-loss, but with only $\sim 20$ subhaloes per bin, the statistics are weak and subject to orbit dependent variations. In contrast, our lowest mass bins have $\sim 200$ haloes per bin. For satellites with $d_{\rm{peri}} \le 30 \, \mathrm{kpc}$, no SIDM satellites last longer than $6 \, \mathrm{Gyr}$, whereas some CDM satellites remain throughout the duration of accretion activity for nearly $11 \, \mathrm{Gyr}$.

\begin{figure*}
\centering
\includegraphics[width=0.68\textwidth]{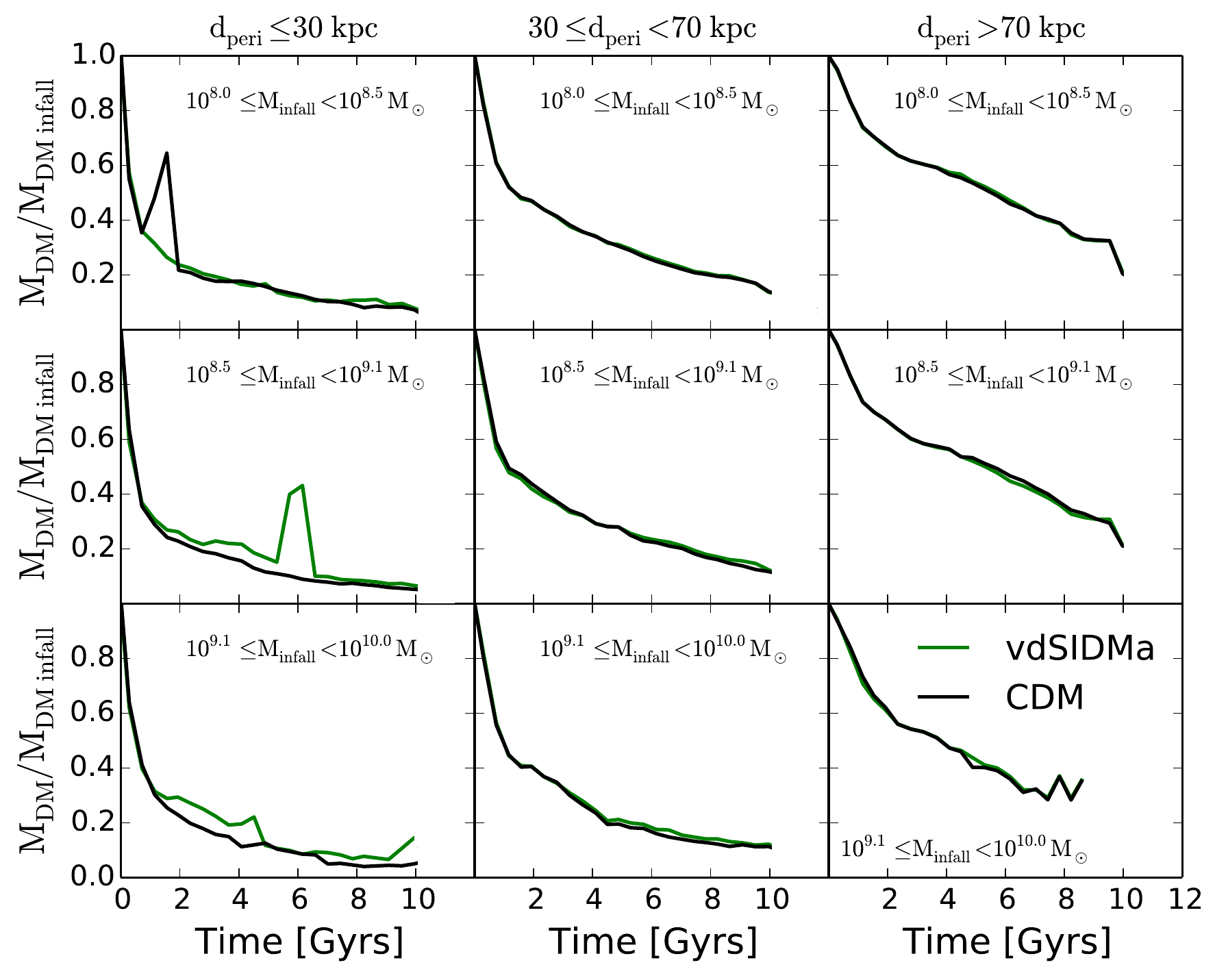}
\caption{Total mass remaining in satellites over time binned by mean pericentre, $d_{\rm{peri}}$, and infall mass, $M_{\rm{infall}}$. Subhalo evaporation is small enough to render no observable difference between vdSIDMa and CDM. Simulation vdSIDMb (not shown) shows identical results with no difference with CDM.}
\label{fig:allmass_strip_v_time_p2}
\end{figure*}

\begin{figure*}
\centering
\includegraphics[width=0.68\textwidth]{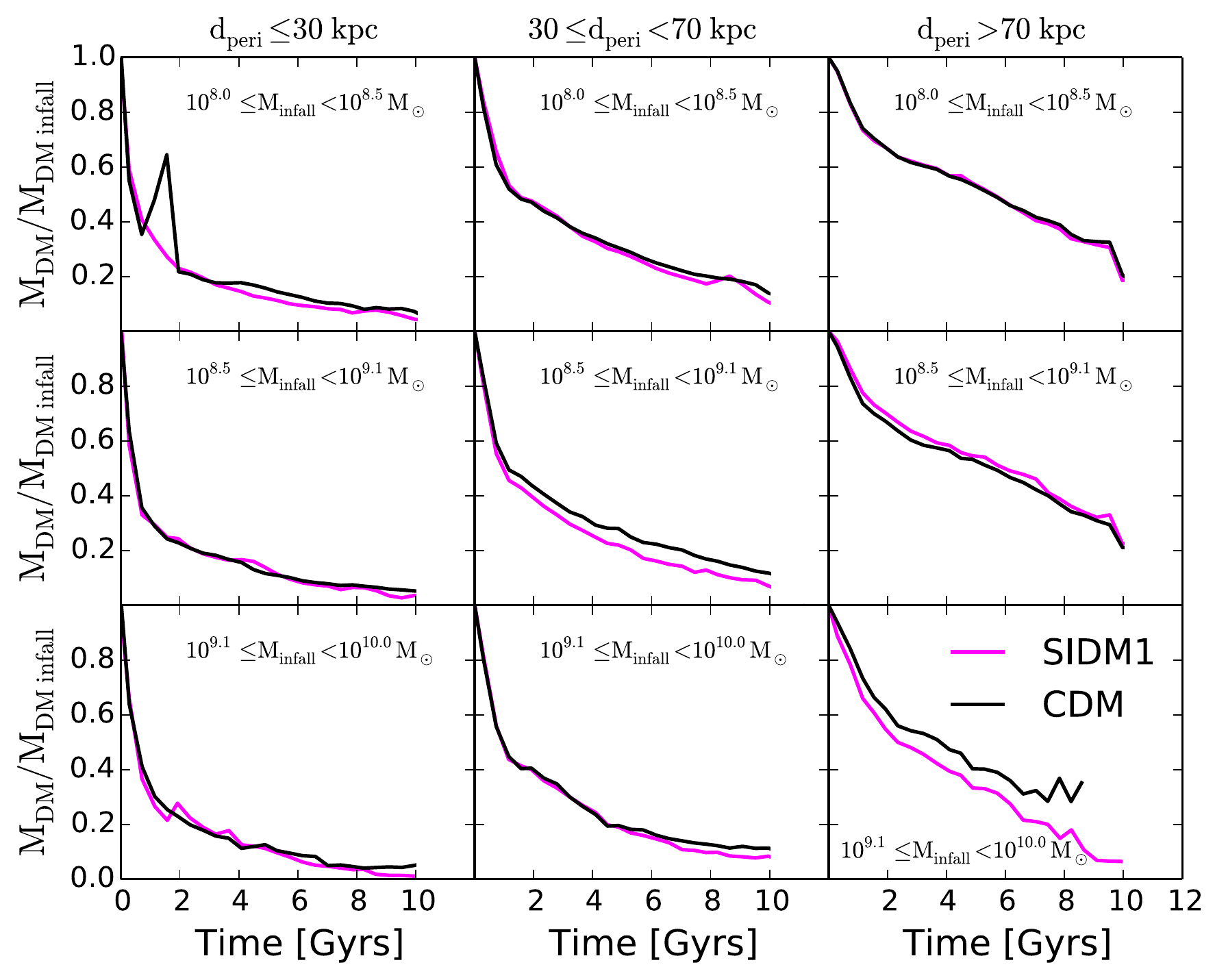}
\caption{Total mass remaining in satellites over time binned by mean pericentre, $d_{\rm{peri}}$, and infall mass, $M_{\rm{infall}}$. Subhalo evaporation has a very weak effect in increasing the rate of mass-loss in SIDM1 relative to CDM. The extra mass-loss seen in the highest mass and largest pericentre bin arises mostly due to different satellite orbits of a limited sample, and not due to subhalo evaporation.}
\label{fig:allmass_strip_v_time_p4}
\end{figure*}

Fig.~\ref{fig:allmass_strip_v_time_p2} shows results for vdSIDMa. Aside from halo finding noise at low pericentres, the SIDM results are strikingly similar to those of CDM. Similarly, nothing different happens in the case of vdSIDMb. This indicates that subhalo evaporation is not significant for the v-d models, and also that tidal stripping does not affect the DM loss rate differently in cored versus cuspy subhaloes in the regime presented. Only for subhaloes with low pericentre will a core cause increased tidal stripping in the total DM mass, as discussed in Fig.~\ref{fig:tidal_radius}. Without the added tidal forces of a disc in the host (see Section~\ref{sec:disc}) and with few subhaloes ever achieving very low pericentre, $r_{\rm{peri}} < 10 \, \mathrm{kpc}$, the average mass-loss over time is not substantially altered by differential tidal stripping in cored and cuspy subhaloes. Lastly, in Fig.~\ref{fig:allmass_strip_v_time_p4}, SIDM1 shows hints of a weak effect in some instances, with stripping increased relative to CDM by just a few percent after $10 \, \mathrm{Gyr}$ in orbit. However, as with SIDM10, the highest mass bins have a limited sample of haloes and orbits are perturbed relative to CDM more in the v-i models than in the v-d models. In particular, the extra mass-loss seen in the highest mass and largest pericentre bin of SIDM1 arises mostly due to different satellite orbits, not because of subhalo evaporation.

\begin{figure}
\includegraphics[width=0.48\textwidth]{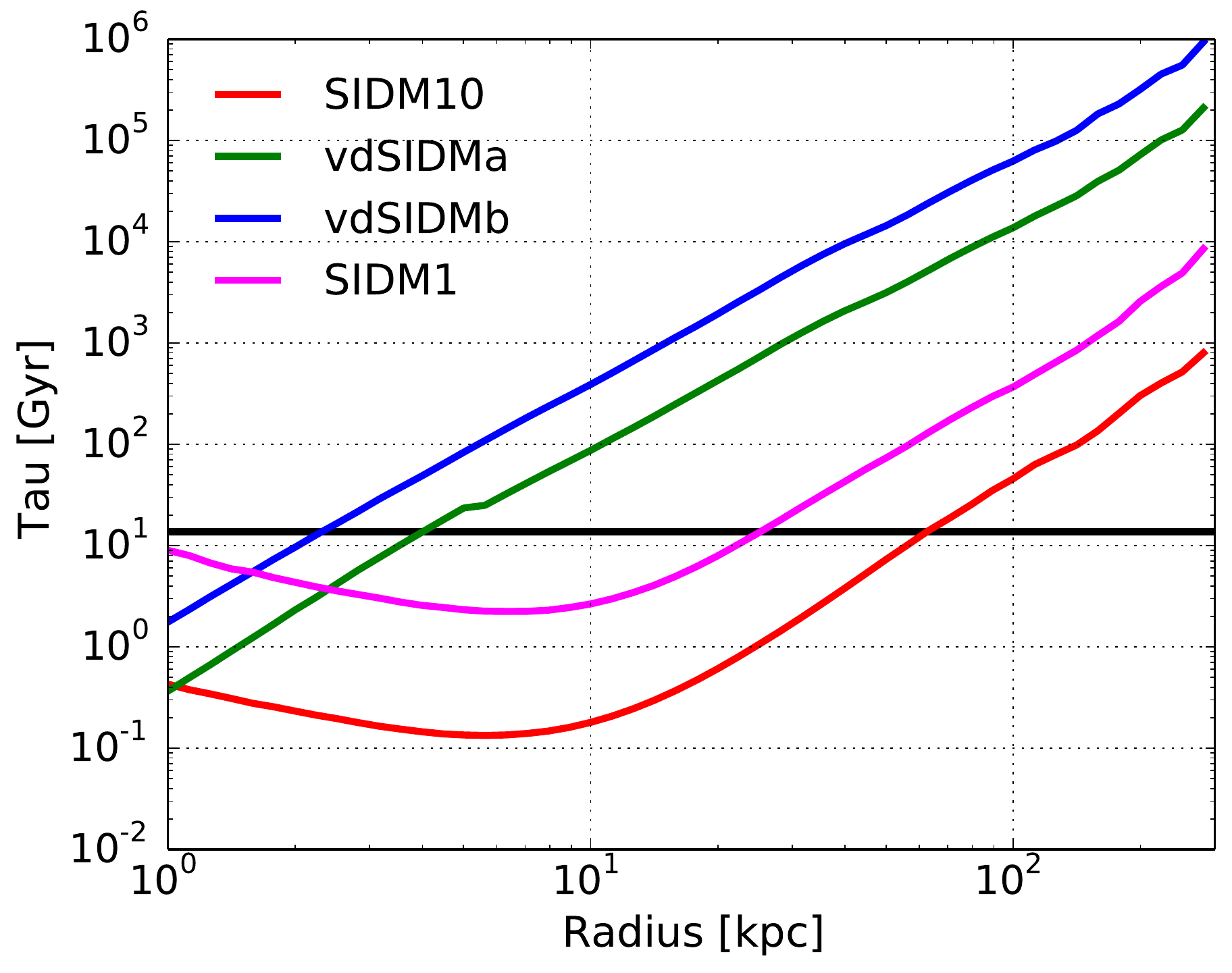}
\caption{Mean time between collisions between background host halo particles and particles within a subhalo on a circular orbit as a function of orbital radius. The black horizontal line is the age of the universe.}
\label{fig:timescales}
\end{figure}

These results all fit expectations when considering the typical time-scale for a subhalo particle to collide with the background under different orbits. Using the scattering rate of equation~(\ref{eq:scattering}), cross-sections shown in Fig.~\ref{fig:cross_sections}, and assuming constant orbital velocity, $v_{\rm{orbit}}$, the mean time between collisions for a given particle, $\tau$, is given by
\begin{equation}
\tau = \left( \rho(r) \frac{\sigma(v_{\rm{orbit}})}{m_\chi} v_{\rm{orbit}} \right)^{-1}
\end{equation}
This time-scale is shown in Fig.~\ref{fig:timescales} for each of our SIDM models as a function of radius for circular orbits. The value of $\rho(r)$ is taken from each simulation's respective host at $z=0$, averaging out spherical asymmetries, and used in turn to compute $v_{\rm{orbit}}$. Since we do not use an evolving host potential, yet host haloes grow in time, $\tau$ is underpredicted for a true dynamical system. We draw a line at $t_{\rm{Hubble}} = 13.7 \, \mathrm{Gyr}$ to highlight the maximum length of time a satellite can spend in a Milky Way-like system and experience evaporation. Subhalo evaporation becomes important for all radii where $\tau < t_{\rm{Hubble}}$.

For SIDM10, $\tau < t_{\rm{Hubble}}$ for all orbits within $60 \, \mathrm{kpc}$ of the host. Many subhaloes spend a large proportion of their time within this distance, meaning evaporation should noticeably increase the mass-loss rate. SIDM1 has the next strongest level of interaction, a full order of magnitude weaker. Here $\tau = t_{\rm{Hubble}}$ at a radius of $25 \, \mathrm{kpc}$. Haloes that spend a Hubble time within this distance will be fully disrupted due to tidal forces and thus not exist today. Yet the radius is still large enough for satellites to pass within during pericentre and survive with a few percent increase in mass-loss. The v-d models have such long time-scales for scattering that this simple time-scales argument suggests that subhalo evaporation should not have an observable effect on total mass-loss. All three of these predictions agree with the results of Figs~\ref{fig:allmass_strip_v_time_p1}-\ref{fig:allmass_strip_v_time_p4}. Interestingly, due to the host halo density flattening out in the core of our v-i simulations, $\tau$ plateaus and even rises as $R$ decreases below $10 \, \mathrm{kpc}$.

Since subhalo evaporation does not substantially affect satellites in our viable SIDM models, we rule out the possibility of evaporation contributing to the enhanced loss of stars. Additionally, the lack of change in total mass-loss rate in the v-d models also indicates that tidal stripping does not affect cored SIDM subhaloes more than CDM subhaloes with regard to total DM mass-loss. Instead, another mechanism involving an outward migration of stars must exist. Since tidal stripping preferentially removes mass from the outer halo, but stars originate in the inner halo, any outward transport of stars can lead to increased stellar mass-loss without a simultaneous increase in total mass-loss. We discuss this mechanism in the next section.

\section{Analytic Explanation for Enhanced Stellar Stripping}
\label{sec:theory}
To explain the results of Section \ref{sec:data} without subhalo evaporation, we turn to the fundamental physics of tidal stripping. External to subhaloes, the strength of the gravitational tidal field controls how much mass is stripped from satellites. This plays a role in determining the tidal radius, $r_{\rm{t}}$, which defines an approximate distance beyond which mass is no longer bound to a subhalo. Naturally, a smaller tidal radius results in more stripping. Internal to subhaloes, which stellar particles are stripped is controlled by dynamics at the centre of the halo. To compare CDM against SIDM, the size of the halo core, $r_{\rm{core}}$, is critical since it correlates with the strength of dynamical differences due to self-interactions. SIDM reduces halo central density and increases halo central velocity dispersion, reducing the binding energy of particles within the core. Consequently, an increase in stellar mass-loss manifests in two ways. 

\begin{figure*}
\centering
\includegraphics[width=0.68\textwidth]{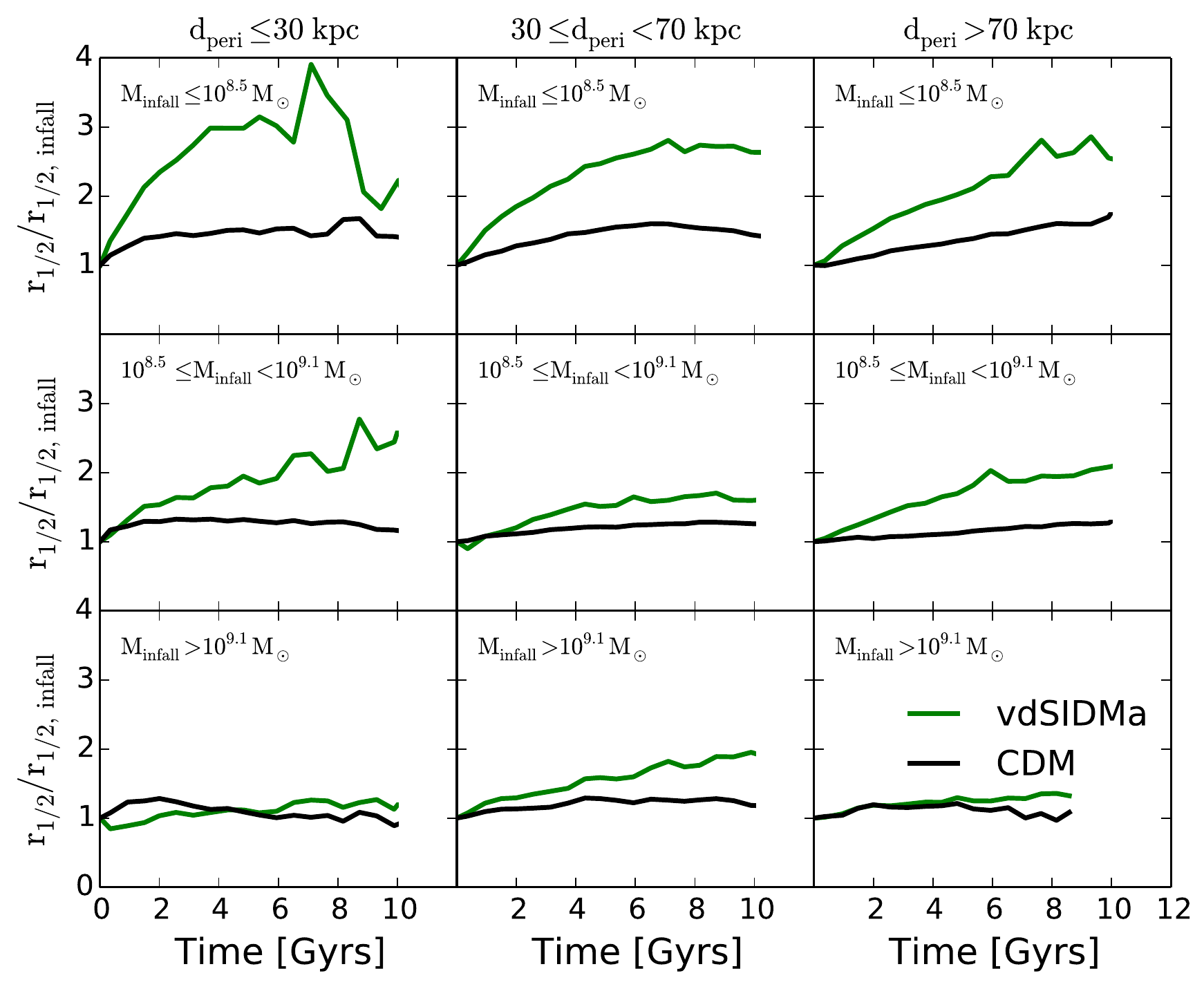}
\caption{The ratio of the half-mass radius of stars that remain bound to a satellite as a function of time to the half-mass radius at infall. Satellites are binned into groups of different mean orbital pericentre and infall mass, and their distributions averaged per bin. Stars in SIDM disperse out to larger radii than in CDM, starting immediately after tagging. This is a primary driver of the increase in stellar mass stripping.}
\label{fig:star_tag_dist}
\end{figure*}

First, stars initially formed in the centre of the subhalo are more likely to disperse out to larger radii, and thus more likely to pass beyond the tidal radius. In a hydrodynamical simulation of a field dwarf galaxy with SIDM, \cite{Vogelsberger14} demonstrated that the distribution of stars within the core is tied to the dark matter distribution, and that the dark matter halo core size grows with time. As a result, stars tend to spread out over time. We show this for our simulations in Fig.~\ref{fig:star_tag_dist}, plotting the ratio of the half-mass radius of the stellar component, $r_{1/2}$, as a function of time to the half-mass radius of the stellar component at infall. Just as in Figs~\ref{fig:strip_v_time_p2}-\ref{fig:strip_v_time_p4}, we take the average of this radius ratio for all haloes belonging to a certain infall mass and mean pericentre bin. In the vdSIDMa case, stars quickly puff out to larger radii whereas in the corresponding CDM case, $r_{1/2}$ grows at a much slower rate over time or remains constant. The relative strength with which SIDM stars spread out in each infall mass and mean pericentre bin is well correlated with the strength of the enhancement of stellar stripping for all SIDM models. As such, we only include the example of vdSIDMa.

\begin{figure}
\includegraphics[width=0.48\textwidth]{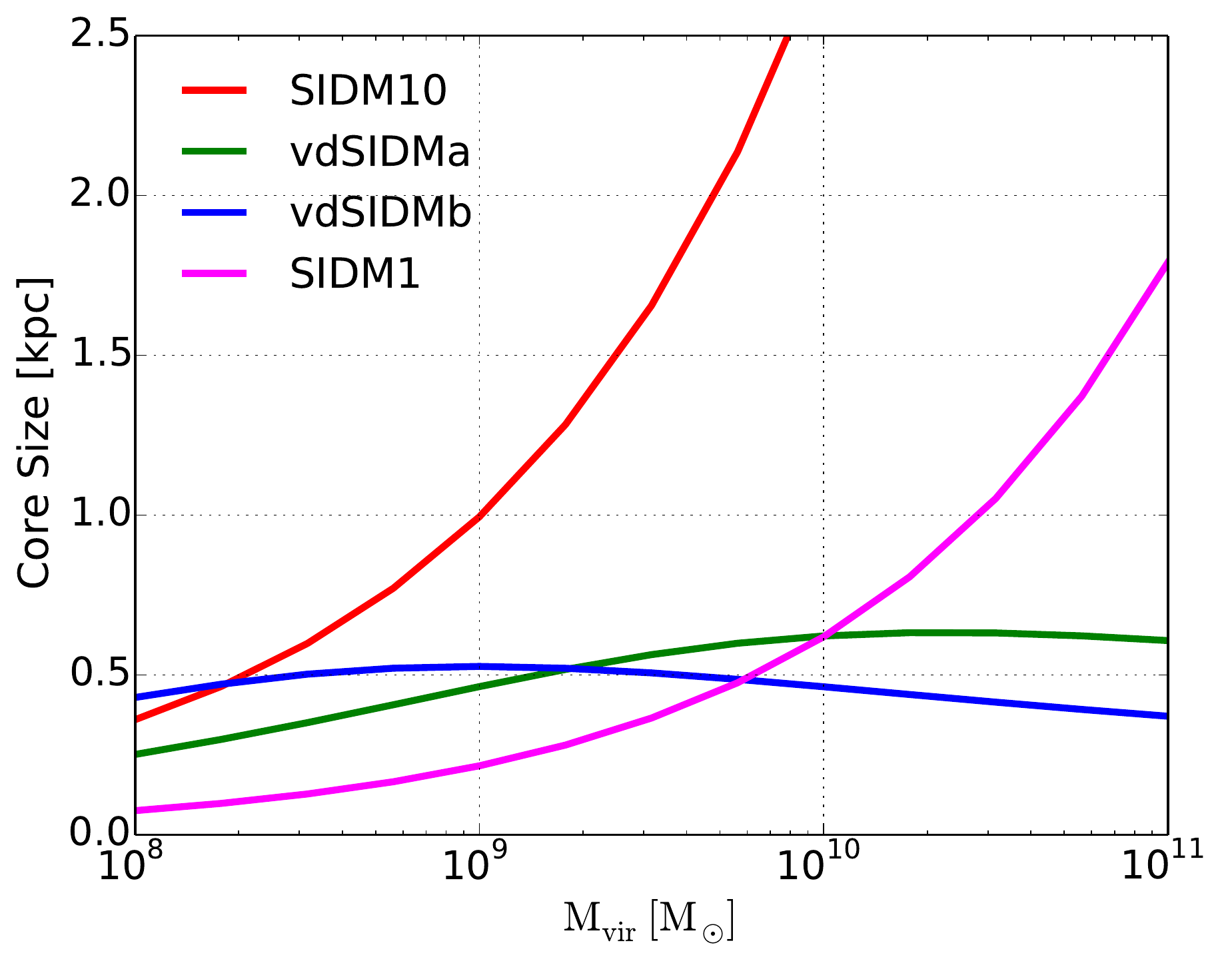}
\caption{Approximate tidal radius of subhaloes in the gravitational field of a $2.2 \times 10^{12} \, \mathrm{M_{\sun}}$ Hernquist profile CDM-like host at distances of $30, 15$, and $5 \, \mathrm{kpc}$. Shown in blue is the tidal radius for subhaloes with cores of the size predicted for vdSIDMb. At low orbital distances and low halo masses, the tidal radius begins to lie within the low density core, resulting in a reduction in $r_{\rm{t}}$ compared to cuspy haloes (solid black lines). If the host halo includes a disc potential (bottom panel), tidal forces are increased at low radii and $r_{\rm{t}}$ is reduced even further compared to cuspy haloes. We assume a disc of mass $1 \times 10^{11} \, \mathrm{M_{\sun}}$ of the same form as used in \protect\cite{Penarrubia10}. For orbital distances larger than $30 \, \mathrm{kpc}$ in all SIDM models, $r_{\rm{t}}$ is negligibly affected by cores in subhaloes.}

\label{fig:tidal_radius}
\end{figure}

Secondly, for a fixed subhalo mass and tidal field, the tidal radius is diminished in cored haloes for orbits where $r_{\rm{t}} \le r_{\rm{core}}$. Since the tidal radius lies within the core only for strong tidal fields at low galactocentric radii, the second effect is applicable only for a small fraction of satellites. We highlight this effect in Fig.~\ref{fig:tidal_radius}, showing the tidal radius of a satellite at $30$, $15$, and $5 \, \mathrm{kpc}$ from the host halo centre for a cored SIDM and a cuspy CDM subhalo.

With $r_{\rm{t}}$ and $r_{\rm{core}}$ as controlling factors, we postulate that the relative strength of stellar mass stripping will follow the ratio $r_{\rm{core}}/r_{\rm{t}}$, an indicator of how close stripping is to where stars reside. Using simple models, we compute this ratio as a function of subhalo mass and dark matter model.

As discussed in \cite{Rocha13}, the radius of the cores in SIDM satellites can be estimated by the radius at which particles are expected to scatter once per Hubble time, or approximately once per $10$ Gyr halo lifetime since haloes first form after a few Gyr. They find this predicts the magnitude of core sizes well for the equivalent of our SIDM1 simulation. However, \cite{Rocha13} fit haloes with a Burkert profile \citep{Burkert95} and use the Burkert scale radius, $r_{\rm{b}}$, to define the core size for SIDM1, commenting that it is likely a lucky coincidence that this works for SIDM1, and would not work as well for other models. We instead choose to define core sizes as the radius at which the density profile of an SIDM satellite diverges from that of a CDM satellite with otherwise equivalent conditions. This definition then accurately probes where binding energies will differ due to SIDM.

By matching isolated field haloes from the same initial overdensity of SIDM with those in CDM, we find the SIDM density profile typically breaks away from that of CDM when the slope of the log-log profile is
\begin{equation}
\rm{d} \ln{\rho(r)}/\rm{d} \ln{r} = -1.
\end{equation}
Our simulation's core sizes measured in this manner best align with the purely analytic estimate when we choose the core size as the radius at which particles scatter once per Gyr instead of once per 10 Gyr. Each of these two methods agree roughly in magnitude, and agree with the scaling of $r_{\rm{core}}$ with halo mass. Thus we define $r_{\rm{core}}$ as the distance where
\begin{equation} \label{eq:core_size}
\frac{\rho(r_{\rm{core}})}{m_\chi} \langle \sigma v \rangle (r_{\rm{core}}) = \mathrm{1 \, Gyr^{-1}}.
\end{equation}
The value $\langle \sigma v \rangle(r)$ is the local thermal average of the transfer cross-section times the relative velocity, estimated as in \cite{Vogelsberger12} by an average over a Maxwell Boltzman distribution function:
\begin{equation}
\langle \sigma v \rangle (r) = \frac{1}{2 \sigma_{\rm{vel}}^3(r)\sqrt{\pi}} \int \ (\sigma v)v^2 \rm{e}^{-v^2/4 \sigma^2_{\rm{vel}}(r) \ \rm{d}v},
\end{equation}
where $\sigma_{\rm{vel}}(r)$ is the local velocity dispersion.

Following \cite{Vogelsberger12}, we model our haloes as Hernquist profiles \citep{Hernquist90} with density varying as
\begin{equation} \label{eq:hernquist_rho}
\rho(r) = \frac{Ma}{2\rm{\pi} r} \frac{1}{(r+a)^3}
\end{equation}
and velocity dispersion $\sigma^2_{\rm{vel}}(r,a,M)$ taken directly from \cite{Hernquist90}.
$M$ is an indication of the halo mass, and $a$ is the scale radius. Using our CDM simulation and definition of virial radius, we find that the halo concentration $c \equiv r_{\rm{vir}}/a$ varies as $c = 80.1 \times \left(\frac{M}{\mathrm{M_{\sun}}} \right)^{-1/11}$ at $z=0$. The value of $M$ in terms of our virial mass, $m_{\rm{vir}}$ is computed as
\begin{equation}
M = m_{\rm{vir}} \frac{(1+c)^2}{c^2}
\end{equation}
and the scale radius is simply $a = r_{\rm{vir}}/c$.

\begin{figure}
\includegraphics[width=0.48\textwidth]{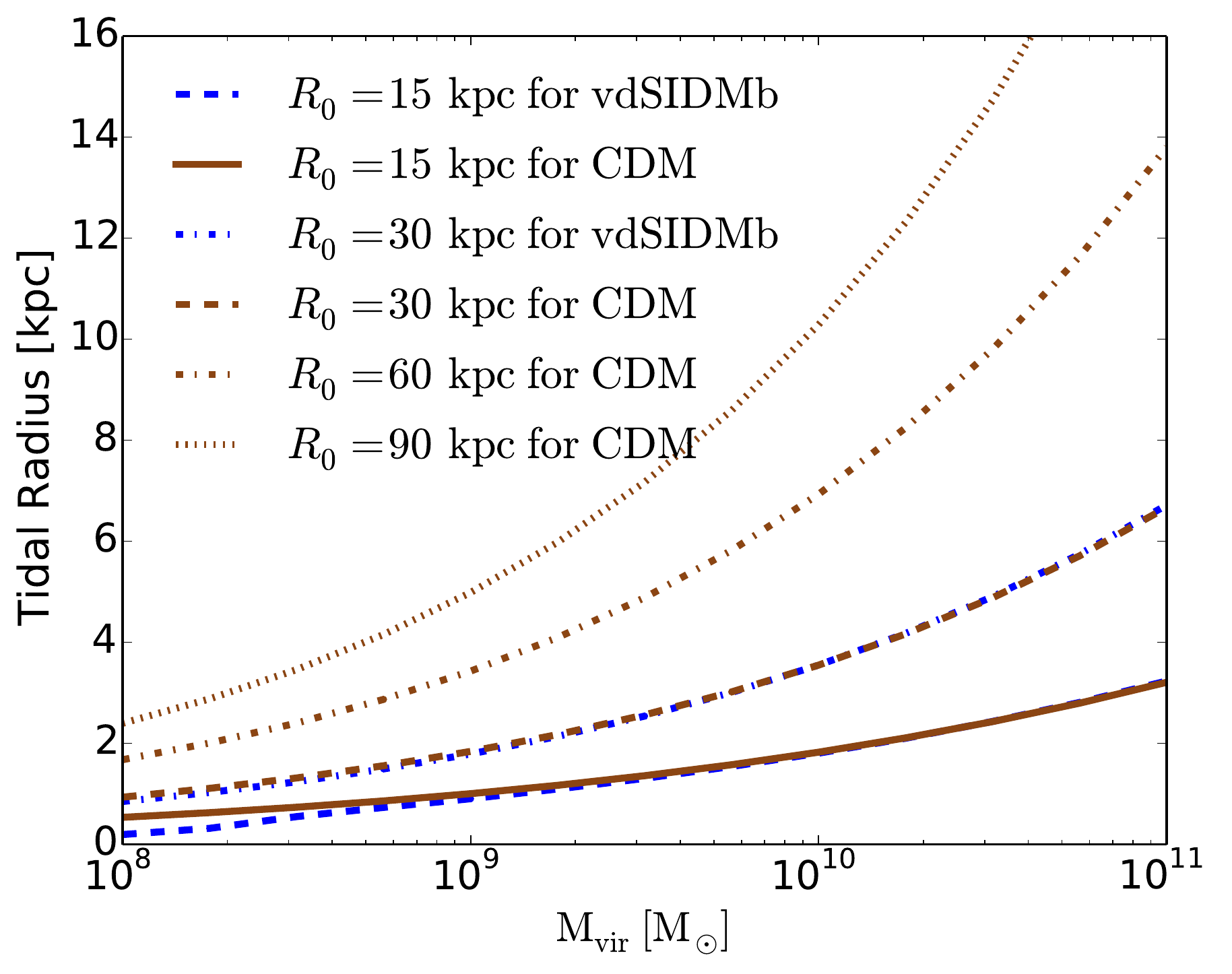}
\caption{Core sizes as a function of virial mass predicted for haloes in each SIDM model. Core size is estimated as the radius where particles scatter on average once per Gyr for an isolated halo.}
\label{fig:analytic_core_sizes}
\end{figure}

Now for any $m_{\rm{vir}}$ we use equation~(\ref{eq:core_size}) to compute the core sizes of each of our SIDM models, as shown in Fig~\ref{fig:analytic_core_sizes}. In the two velocity-independent models, the core size scales with halo mass. As central density increases, scattering increases. In the velocity-dependent models, there is a competing effect of higher densities and higher velocity dispersions (and thus lower cross-section, see Fig.~\ref{fig:cross_sections}) making the core size vary only slightly with halo mass. Consequently, core sizes of vdSIDMa and vdSIDMb are lower than that of SIDM1 for high mass haloes, then overtake SIDM1 between $10^9$ and $10^{10} \, \mathrm{M_{\sun}}$ going towards lower halo mass. Due to its cross-section's steep dependence on velocity, vdSIDMb even overtakes SIDM10 for haloes smaller than $10^8 \, \mathrm{M_{\sun}}$ in spite of having the smallest cores above $10^{10} \ \mathrm{M_{\sun}}$. 

To continue our estimate of $r_{\rm{core}}/r_{\rm{t}}$, we need to compute the tidal radius. In a three body system of host, satellite, and star, the last closed effective potential surface indicates the region in which stars will still be bound to the subhalo. The distance to this surface at the $L_3$ Lagrange point is a good approximation of $r_t$. For a circular orbit and spherically symmetric host and subhalo, $r_{\rm{t}}$ is given in \cite{BinneyTremaine} and rewritten here as 
\begin{equation} \label{eq:tidal_radius}
r_{\rm{t}} = R_0 \left( \frac{m(r_{\rm{t}})}{M(R_0)(3-\frac{d \ln{M}}{d \ln{R}}\rvert_{R=R_0} )}  \right)^{1/3},
\end{equation}
where $R_0$ is the radius of the satellite's orbit, $m(r_{\rm{t}})$ is the mass enclosed by the satellite at a distance $r_{\rm{t}}$ from its centre, and $M(R)$ is the mass enclosed by the host at distance $R$ from the host's centre.

Isolating how the tidal radius varies with satellite mass, we compute and plot in Fig.~\ref{fig:tidal_radius} the tidal radius for subhaloes within a $2.2 \times 10^{12} \, \mathrm{M_{\sun}}$ CDM-like host at various fixed distances to the host halo's centre. The tidal radius increases with both satellite mass, $m_{\rm{sat}}$, and galactocentric distance, $R_0$, closely following the approximate equation 
\begin{equation}
r_{\rm{t}} \approx 0.7 \left(\frac{m_{\rm{sat}}}{M_{host}} \right)^{1/3} R_0
\end{equation}
for Hernquist haloes. Since the tidal radius depends on the mass enclosed within a satellite at a given radius, and the mass enclosed is reduced within a core, the tidal radius is suppressed for SIDM haloes when $r_{\rm{t}} \le r_{\rm{core}}$. Just beyond the core radius, there is a buildup of excess mass in SIDM haloes relative to their equivalent CDM haloes since mass displaced from the halo centre is not removed entirely. As a result, the same total mass is enclosed at all larger radii, and there is no difference in $r_{\rm{t}}$ for SIDM and CDM haloes. We verify this effect by modifying the Hernquist profile for our satellites to enforce $\rm{d} \ln{\rho(r)}/\rm{d} \ln{r} = -0.3$ within $r_{\rm{core}}$ while preserving the same total virial mass. While a true core has a logarithmic density profile slope of $0.0$, the profile of satellites in our simulations is rarely completely flat and instead closer to $-0.3$.

As seen in the blue dashed lines of Fig.~\ref{fig:tidal_radius}, the tidal radius of cored satellites is reduced relative to that of a cuspy halo (black solid lines) for orbits where $R_0 \lesssim 30 \, \mathrm{kpc}$, and the reduction becomes increasingly important for lower galactocentric radii. The effect is also magnified as the satellite core size increases relative to the tidal radius of the same mass cuspy satellite. For the vdSIDMb model illustrated, this happens towards lower satellite masses. The bottom panel of Fig.~\ref{fig:tidal_radius} illustrates how the presence of a disc in the host halo would enhance these effects. Here, we assume a disc of mass $1 \times 10^{11} \, \mathrm{M_{\sun}}$ of the same form as used in \cite{Penarrubia10}, and reduce the host's dark matter mass so the sum total of halo plus disc mass is still $2.2 \times 10^{12} \, \mathrm{M_{\sun}}$. Beyond $30 \, \mathrm{kpc}$ for all SIDM models, the change in $r_{\rm{t}}$ is negligible even with the presence of a disc.

\begin{figure}
\includegraphics[width=0.48\textwidth]{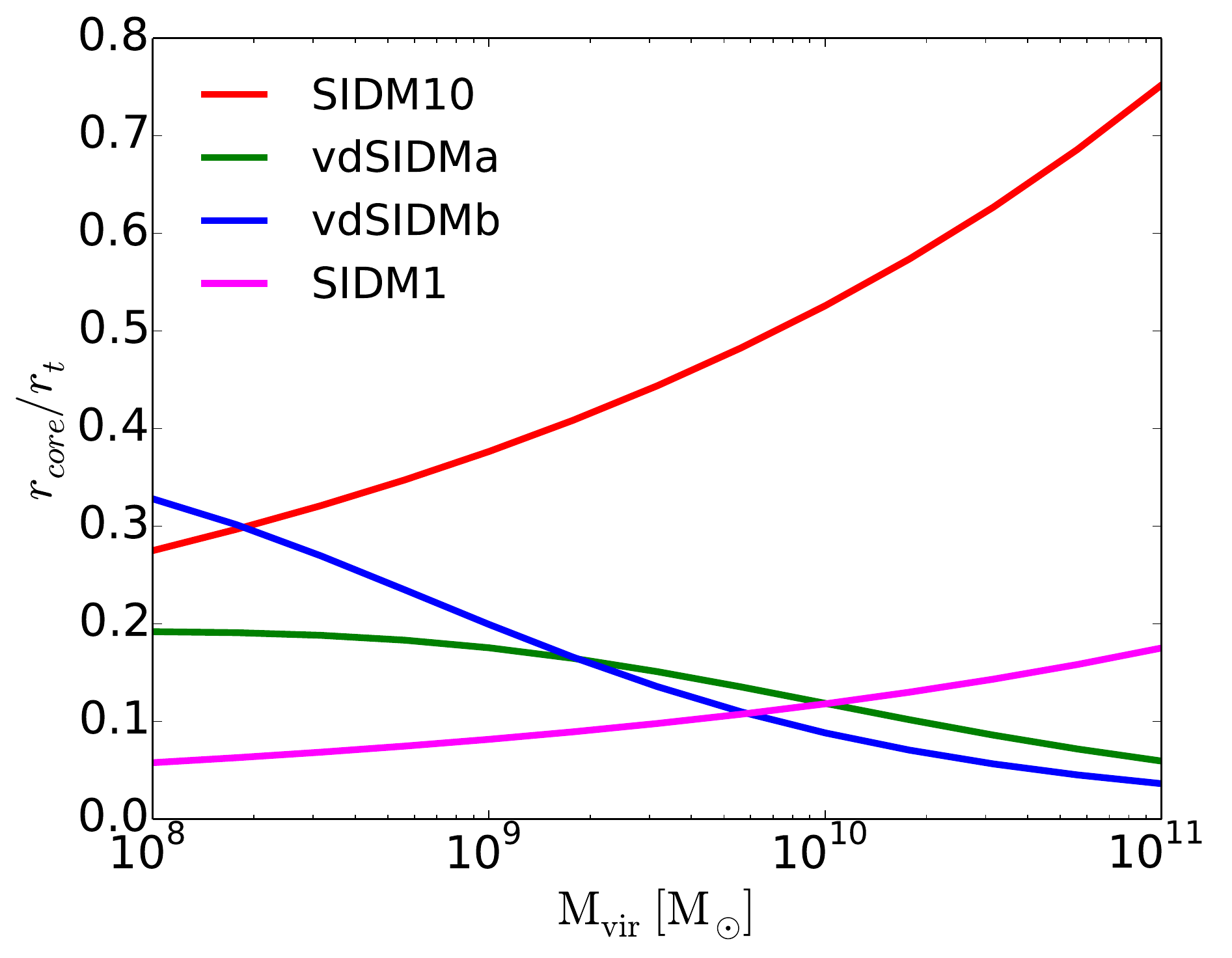}
\caption{Ratio of the core size of subhaloes to tidal radius in a $2.2 \times 10^{12} \, \mathrm{M_{\sun}}$ host with subhaloes placed at a distance of $45$ kpc from the host's centre. Higher values of this ratio indicate a higher propensity for stellar mass stripping. The inverse dependence of cross-section on relative velocity in vdSIDMa and vdSIDMb leads to more enhanced stripping in low-mass satellites, while the opposite is true in constant cross-section models.}
\label{fig:core_over_tidal}
\end{figure}

In Fig.~\ref{fig:core_over_tidal}, we now compute the ratio $r_{\rm{core}}/r_{\rm{t}}$ with $r_{\rm{t}}$ found for a satellite at a distance of $45 \, \mathrm{kpc}$ from the host. The ratio shows the intricacies of the effects of each SIDM model and the mass dependence of the tidal radius. In the v-i models, due to  $r_{\rm{core}}/r_{\rm{t}}$ increasing as a function of halo mass, we expect the disparity in stripping relative to CDM to increase with halo mass. Indeed, as shown in Section \ref{sec:data}, SIDM1 does have a higher rate of stripping relative to CDM for larger mass haloes when $d_{\rm{peri}} > 30 \, \mathrm{kpc}$. This is driven by core sizes that increase in size faster than the tidal radius for more massive subhaloes. The opposite is true of the v-d models. Here, core sizes vary only weakly with mass, allowing for larger haloes to more effectively shield their inner region via larger tidal radii. The vdSIDMb model particularly demonstrates this where haloes above $10^{10} \, \mathrm{M_{\sun}}$ should exhibit little difference with stripping in CDM, but haloes with a mass of $10^8 \, \mathrm{M_{\sun}}$ magnify the effects of SIDM enhanced stripping the most of any model, including the high cross section case of SIDM10. This is precisely what the simulation data in Figures \ref{fig:strip_v_time_p2} and \ref{fig:strip_v_time_p3} show. The SIDM enhancement of tidal stripping is large for low mass haloes, and quickly drops to become a small effect for more massive haloes.

By itself, Figure \ref{fig:core_over_tidal} does not explain the different stellar stripping behaviour in SIDM1 from the v-d models for the interval of $d_{\rm{peri}} \le 30 \, \mathrm{kpc}$. The behaviour arises from differences in the host halo potential. In the v-i models, the host halo develops a substantial core, as can be imagined by extrapolating the core sizes of Fig.~\ref{fig:analytic_core_sizes} to a $2.2 \times 10^{12} \ \mathrm{M_{\sun}}$ halo. In SIDM1, this core causes a reduction in the mass enclosed versus radius profile relative to CDM out to $17 \, \mathrm{kpc}$. In the v-d models, the collisional cross-section drops enough in the high velocity dispersion region of the host's centre to affect only $\sim 2 \, \mathrm{kpc}$ in vdSIDMa and $\sim 1 \, \mathrm{kpc}$ in vdSIDMb. Since the strength of the tidal force, as approximated in equation~(\ref{eq:tidal_radius}), depends on the mass enclosed in the host for a fixed radius, haloes in the SIDM1 experience a reduction in $r_{\rm{t}}$ relative to the v-d models when they pass near or within $17 \, \mathrm{kpc}$. Since haloes have extended distributions, a subhalo need only pass near $17 \, \mathrm{kpc}$ for part of its extremities to pass within the critical radius where tidal forces are reduced relative to CDM. Therefore, any subhalo in SIDM1 orbiting in this range will face competing effects of the subhalo core increasing the likelihood for stars to escape relative to CDM, but also the host halo core reducing tidal forces and decreasing this likelihood. The result seen in the first two rows of column 1 in Fig.~\ref{fig:strip_v_time_p4} suggests that these effects cancel each other out enough to produce a CDM level of stripping. In the case of the highest mass bin at low pericentre, low number statistics lead to more chaotic behaviour.

\section{Implications}
\label{sec:implications}
In this section, we highlight the consequences of SIDM on the $z=0$ ensemble of satellites, and the stellar halo.

\subsection{Mass Functions}
As demonstrated in Section \ref{sec:data}, SIDM causes an increase in the mass-loss rate of stellar material but does not change significantly the mass-loss rate of the total dark matter mass. In turn, the subhalo mass function in terms of dark matter is not affected, but the stellar mass function should be suppressed. Due to limitations of not tagging the same total number of haloes in SIDM as in CDM, however, we cannot directly show the suppression.

\begin{figure}
\includegraphics[width=0.48\textwidth]{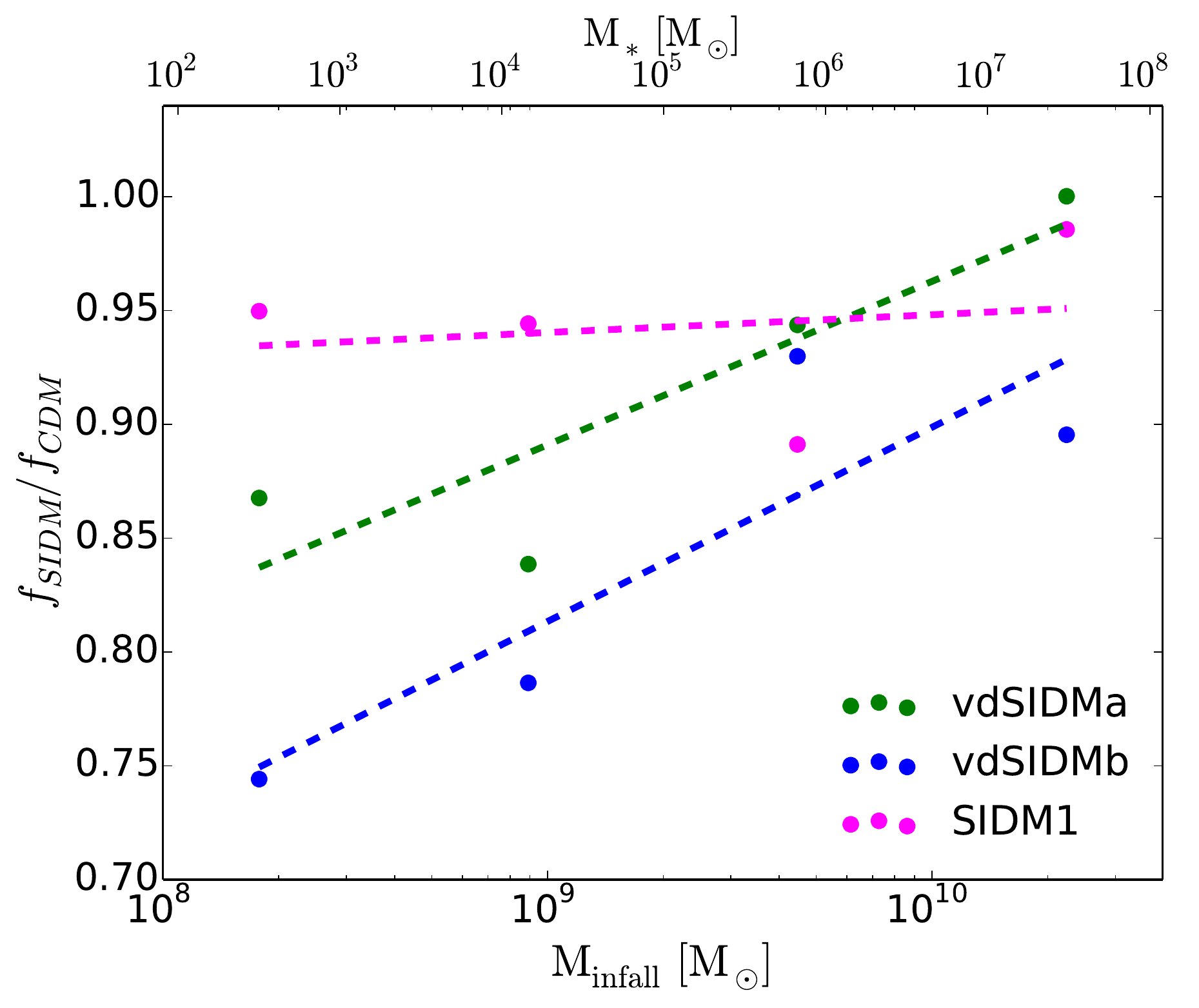}
\caption{Ratio of the fraction of stars remaining in SIDM haloes to that in CDM haloes at $z=0$ as a function of halo infall mass. SIDM haloes lose more stars, which in turn causes a suppression in the stellar mass function. The strength of this effect correlates well with the size of the core radius relative to the tidal radius as a function of mass (see Section~\ref{sec:theory}). The upper axis shows the approximate stellar mass per satellite based on \protect\cite{Moster13} abundance matching.}
\label{fig:ratios_z0}
\end{figure}

Instead, in Fig.~\ref{fig:ratios_z0} we show the fraction of stars stripped since infall for $z=0$ satellites in SIDM, $f_{\rm{SIDM}}$, divided by the fraction of stars stripped since infall for $z=0$ satellites in CDM, $f_{\rm{CDM}}$, versus the infall mass of satellites. For reference, we include the approximate stellar mass per satellite on the upper axis based on \cite{Moster13} abundance matching. The signals seen for all satellites over the entire accretion history of the host persist at the single snapshot of the present day. In all three SIDM models, there is a clear increase in stellar mass stripped relative to CDM. In the v-d models, the velocity dependence of the cross-section shows its effects clearly. Lower mass satellites with larger cross-sections lead to increased core sizes relative to their subhalo radius and a stronger enhancement of stripping. The effect of SIDM1 is more constant with mass. While these trends do appear, there is also tremendous scatter from halo to halo arising from different infall times and orbits. 

In spite of nearly all stripping occurring while satellites are within $50 \, \mathrm{kpc}$ of the host, there is not a strong correlation of $f_{\rm{SIDM}}/f_{\rm{CDM}}$ with the $z=0$ galactocentric distance of satellites all the way out to $300 \, \mathrm{kpc}$. Beyond that, satellites in both SIDM and CDM are stripped by just a few percent, if at all. Orbits are elliptical enough to disperse affected satellites out to large radii. We caution that due to the nature of our particle tagging, we cannot definitely declare the magnitude of the signal, only the trends. A full hydrodynamical simulation, or careful injection of high resolution, non-scattering star particles would be needed to establish the magnitude properly.

\subsection{Stellar half-mass radius}

\begin{figure}
\includegraphics[width=0.48\textwidth]{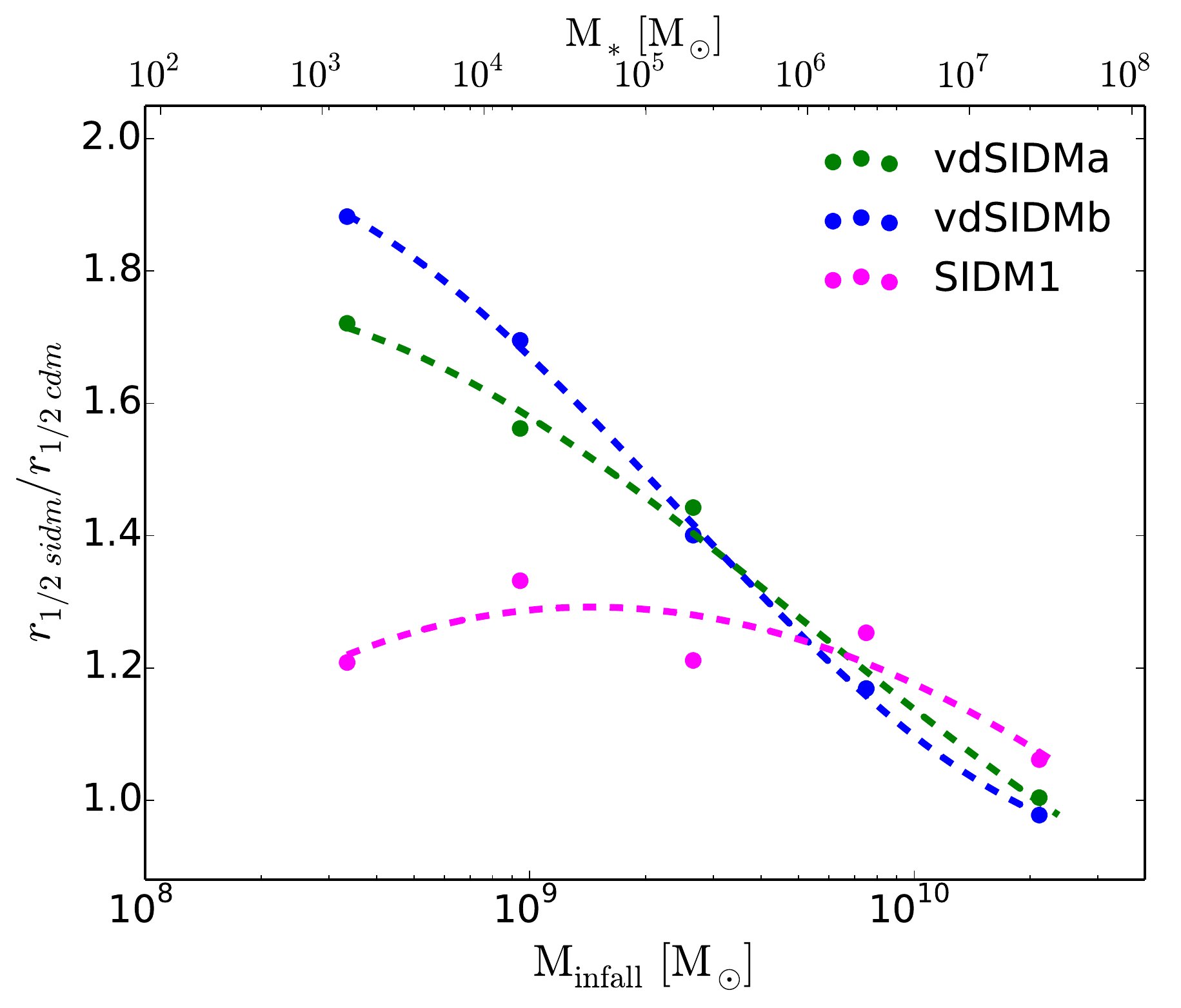}
\caption{SIDM to CDM ratio of the half-mass radius of stars in all satellites at $z=0$ as a function of their infall halo mass. The half-mass radius increases relative to that of CDM in all SIDM models due to stars drifting outwards as satellite cores grow. The upper axis shows the approximate stellar mass per satellite based on \protect\cite{Moster13} abundance matching.}
\label{fig:half_mass_radius}
\end{figure}

In Section, \ref{sec:theory} we demonstrated that stars migrate to larger distances on average in SIDM than in CDM. This results in an increased half-light radius, $r_{1/2}$, an effect also seen for cored haloes in \cite{Vogelsberger14} and \cite{Errani15}. We therefore measure the half-mass radius, a proxy for half-light radius, of all satellites in our simulations at $z=0$. We then take the ratio of the half-mass radius for SIDM to that of CDM and plot it as a function of the subhalo infall mass in Fig.~\ref{fig:half_mass_radius}. Once again we see the same ordering of effects for each of our SIDM models. In all cases, SIDM leads to a larger $r_{1/2}$. For masses above $10^{10} \, \mathrm{M_{\sun}}$, SIDM1 has the largest effect, then vdSIDMa, then vdSIDMb. Towards lower halo masses, the effect increases rapidly for the v-d models. We caution that the magnitude of the increase relative to CDM is not independent of our tagging technique. The trends and relative effect of each model however, reinforced visually by the dashed lines, are much more robust. As in Fig.~\ref{fig:ratios_z0}, we include the approximate stellar mass per satellite on the upper axis based on \cite{Moster13} abundance matching.

\subsection{Stellar Halo}
\label{sec:stellar_halo}

The stellar halo consists of all stars beyond the influence of the host halo's disc. Since our simulations only trace stars formed in accreting satellites, we focus our predictions on the outer halo where accreted stars dominate over stars formed within the host halo. \cite{Pillepich15} find that beyond $\sim 30 \, \mathrm{kpc}$, more than $95\%$ of stellar mass comes from accreted stars in a Milky Way-like host. We therefore consider the stellar halo as the region between $30 \, \mathrm{kpc}$ and $r_{\rm{vir}}$ of our hosts. 

\begin{figure}
\includegraphics[width=0.48\textwidth]{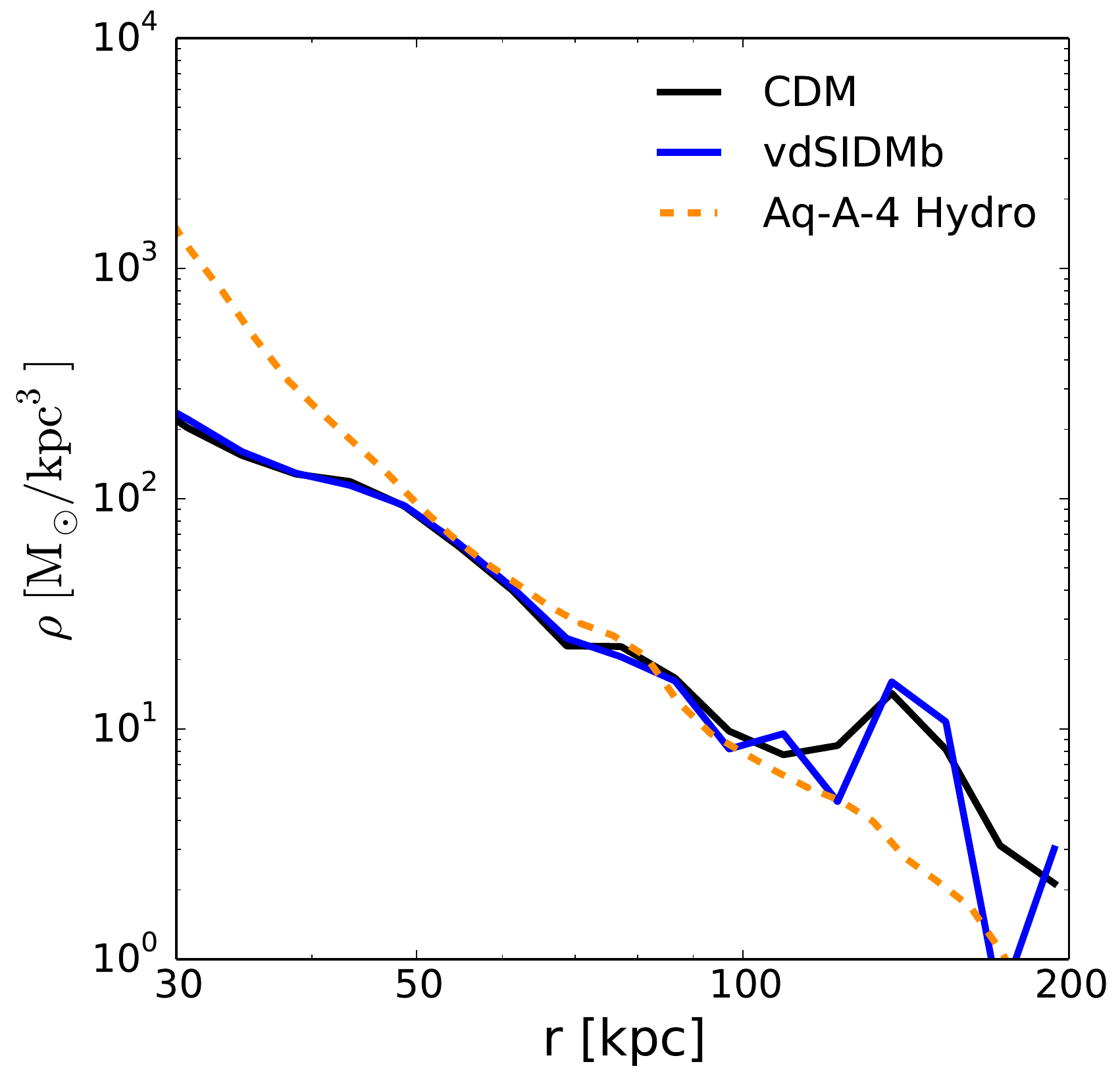}
\caption{Stellar density profile of tagged particles in our host halo due to satellite accretion for our vdSIDMb and corresponding CDM simulations in the outer halo. For comparison we show the stellar density profile of the hydro Aquarius run \protect\citep{Marinacci14} when scaled down in magnitude by a factor of $11$ to match the \protect\cite{Moster13} stellar mass-halo mass relationship. Our tagging technique reproduces the slope very well between $50$ and $100$ kpc where effects of the disc are unimportant. We do not see differences in the stellar density profile between CDM and SIDM in of our simulations.}
\label{fig:stellar_density}
\end{figure}

One major result is the lack of an effect the v-d models have on the stellar halo. In Fig.~\ref{fig:stellar_density}, we show the stellar density profile of vdSIDMb compared to the corresponding CDM case. The two profiles are remarkably similar, with no apparent effect on the density profile due to SIDM. We find the same result for the radial velocity dispersion profile of stars (not shown). To verify that our particle tagging properly reflects features of the stellar halo, we compare the $z=0$ stellar density profile of our host halo to that of the \cite{Marinacci14} simulation, which is a hydrodynamic resimulation of the same initial conditions we used but at a lower DM particle resolution, Aq-A-4. From looking at their $z=0$ field haloes, we find that the slope of their stellar mass-halo mass function is consistent with that of \cite{Moster13}, but is a factor of $11$ higher in magnitude. After uniformly scaling their stellar masses down by that factor of $11$ to better mimic the \cite{Moster13} relation, our stellar density profile matches reasonably well the hydro run (orange line) between $50$ and $100$ kpc.

The v-d models make such little influence due to their small interaction cross-sections in haloes of mass scales above $10^{10} \, \mathrm{M_{\sun}}$. The host halo forms only a very small core, leading to minimal perturbations of the orbits of satellites and the spherically averaged DM dynamics of the host halo in the outer regions. As discussed in Sections \ref{sec:subhalo_evaporation} and \ref{sec:theory}, satellites with $M_{\rm{infall}} > 10^{10} \, \mathrm{M_{\sun}}$ have a low $r_{\rm{core}}/r_{\rm{t}}$ ratio, experiencing minimal enhancement in stellar stripping, and have no subhalo evaporation. Yet $90\%$ of the stellar mass accreted on to the host comes from such satellites. Thus the stellar halo is dominated by of the order of $10$ large satellites which are not substantially affected by SIDM, in a host halo potential which is minimally changed, resulting in no global changes to the stellar halo.

\begin{figure}
\includegraphics[width=0.48\textwidth]{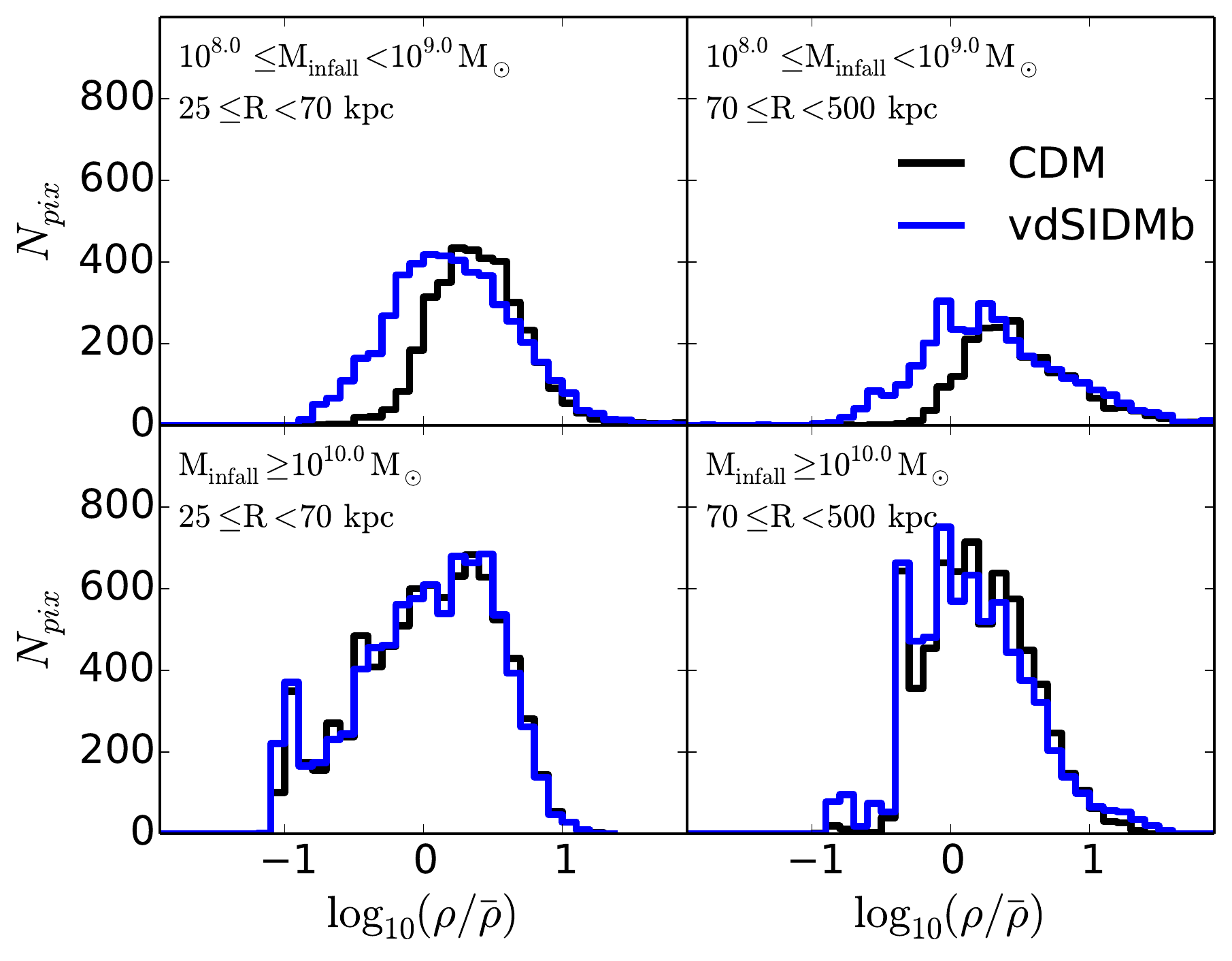}
\caption{Histogram of the density of all stripped stellar mass beyond $30 \, \rm{kpc}$ relative to the mean stellar density, as observed in $3.4$ $\rm{deg}^2$ pixels. Stars are divided into four samples according to the infall mass of the satellite where they came from. We remove stars bound, thus not yet stripped, in subhaloes above $\rm{10^8 M_{\sun}}$ in this sample. Density is given as $\log_{10} (\rho/\bar{\rho})$. Shown is the case of vdSIDMb.}
\label{fig:ratio_hist}
\end{figure}

In order to observe any effects of the v-d models, smaller satellites must be isolated. We highlight one such instance of extracting a signal for vdSIDMb in Fig.~\ref{fig:ratio_hist}. Here, we show a measure of the smoothness of tidally stripped stars across the sky inspired by methods in \cite{Bell08}. We pixelate the sky into $12288$ pixels and find the total amount of stellar mass per pixel that is not bound to an identifiable subhalo. We then compute the ratio of the stellar mass density, $\rho$, to the mean stellar mass density per pixel, $\bar{\rho}$, and take the logarithm to yield $\log_{10}(\rho/\bar{\rho})$. We plot a histogram of these values for stars that came from satellites with infall mass between $10^{8} < M_{\rm{infall}} < 10^{9} \, \mathrm{M_{\sun}}$ (top row), and stars from satellites with infall mass between $10^{10} < M_{\rm{infall}} < 10^{13} \, \mathrm{M_{\sun}}$ (bottom row). Consistent with expectations, no difference exists in the distribution for stars coming from the larger satellites which dominate the total signal. For the lower mass satellites, we do find enhanced stellar stripping. With more stars dispersed across the sky, there is a greater tail of sky patches with a lower than mean density of stars, and fewer sky patches completely devoid of tidally stripped stars ($60$ per cent for vdSIDMb versus $70$ per cent for CDM). 

Both of these distributions hold independent of the radial range of the stellar halo studied. In spite of nearly all stars being stripped from satellites while they are within $50 \, \mathrm{kpc}$ of the host, elliptical orbits carry the effect of enhanced stellar stripping to much larger radii. This is seen in a comparison of the left-hand column which isolates the stellar halo between $25 \le R < 70 \, \mathrm{kpc}$, and the right-hand column which isolates the stellar halo between $70 \le R < 500 \, \mathrm{kpc}$. We caution that this figure serves primarily to illustrate a point and not to find a regime of maximal signal. Different scale sizes of patches on the sky should produce different signals. With limited particle resolution, particularly from stars originating in low-mass satellites, any further investigation is difficult.

In the case of SIDM1, some differences do arise in the stellar density profile and velocity dispersion profile, but primarily result from perturbations of satellite orbits due to a core in the host halo. These perturbations are visually apparent in Fig.~\ref{fig:density_projections}. The largest satellites in SIDM1 do have significant cores and consequently have enhanced stripping relative to CDM for a fixed orbit. They also produce tidal streams with a higher velocity dispersion perpendicular to the orbital plane, as demonstrated in \cite{Errani15} for satellites with cores. However, the global level characteristics of the stellar halo considered are dominated by the particular orbits of the handful of most massive satellites. We are unable to isolate effects directly indicative of SIDM without detailed knowledge of the orbits. On smaller mass scales, SIDM1 does not produce large enough cores as in the v-d models to create significant features like those shown in Fig.~\ref{fig:ratio_hist} for the vdSIDMb model.

\subsection{Alternate Abundance Matching Models}
\label{sec:alternate_abundance_matching}
Due to limited observations, and stochasticity in the stellar mass halo mass relation, abundance matching for haloes with total dark matter mass below $10^{10} \, \mathrm{M_{\sun}}$ is very uncertain. For the sake of assigning approximate stellar masses, we used the \cite{Moster13} abundance matching relationship which influences only Figs~\ref{fig:stellar_density} and \ref{fig:ratio_hist}, and the upper $x$-axis of Figs~\ref{fig:ratios_z0} and \ref{fig:half_mass_radius}. A steeper stellar mass halo mass relationship as suggested in \cite{Brook2014} would mean the largest satellites dominate the stellar halo even more, washing out any signals from v-d models more than already presented. A more shallow relationship, as suggested in \cite{Behroozi13}, would do the opposite. What stellar mass is associated to haloes also affects the luminosity of galaxies in the Milky Way which probe the regimes of influence by each SIDM model. For instance, where v-d models produce the greatest signal, in haloes with infall mass between $10^8$ and $10^9 \, \mathrm{M_{\sun}}$, \cite{Moster13} suggests a stellar mass of $10^2 < M_* < 2 \times 10^4 \, \mathrm{M_{\sun}}$, coinciding with ultrafaint dwarfs. The relationship of Eq. 2 in \cite{Brook2014} extrapolated to lower masses suggests a stellar mass of $2 < M_* < 2 \times 10^3 \, \mathrm{M_{\sun}}$, perhaps too low to be observed. In contrast, the relationship in fig. 4 of \cite{Sawala15} suggests such haloes could host galaxies of approximately $M_* = 10^5 \, \mathrm{M_{\sun}}$, making observations much more tractable.

\subsection{Effect of a Disc}
\label{sec:disc}
A significant effect not modelled in our simulations is the presence of a baryonic disc in the host halo. \cite{Penarrubia10} show that a disc increases the tidal forces on all satellites, and widens the gap in tidal stripping between cuspy and cored satellites. As demonstrated in Fig.~\ref{fig:tidal_radius}, when the tidal radius starts to probe within a satellite core, it drops to lower values with cores than with cusps. Strong tidal forces during low pericentric passages are needed to reach this regime. Since a disc + DM halo increases tidal forces relative to just a DM halo at low radii, it serves to enhance the separation of stellar stripping in cored versus cusped systems. 

We therefore expect all satellites with pericentre $r_{\rm{peri}} \lesssim 30 \, \mathrm{kpc}$ to have more stellar stripping, with SIDM satellites affected more than CDM satellites. Approximately $1/3$ of all existing $z=0$ subhaloes with $m_{\rm{sub}} > 10^8 \, \mathrm{M_{\sun}}$ within the virial radius of $330 \, \mathrm{kpc}$ have passed within $30 \, \mathrm{kpc}$ of the host. To fully quantify the effect, a set of simulations with a disc is needed.

\section{Conclusion}
\label{sec:conclusion}
SIDM may offer a resolution to the too big to fail problem and cusp/core issues of satellite and field galaxies in the Local Group \citep{Vogelsberger12, Zavala13}. Scattering between particles drives mass away from dark matter halo central density cusps, creating lower density cores. Currently, constraints on velocity-independent SIDM places the interesting range of the interaction cross-section at $0.1 \leq \sigma/m_x  \leq 1 \, \mathrm{cm^2/g}$ \citep{Rocha13}. However, the class of velocity-dependent SIDM models, in which the cross-section scales approximately as $\sigma \propto 1/v^\alpha$ for some $\alpha$, remain largely unconstrained. In this paper, we search for new signatures of SIDM on satellite galaxies and their interactions with a host halo to ultimately aid in identifying or constraining SIDM.

We employ a suite of five cosmological DM simulations of Milky Way sized galaxies from \cite{Vogelsberger12}, two with velocity-dependent SIDM models, two with velocity-independent SIDM models, and one with pure CDM. We then carefully tag stellar mass to central particles in satellites at infall, and investigate how SIDM changes their dynamics. Our main results are as follows.

\begin{enumerate}

\item \textbf{SIDM causes an enhanced level of stellar mass-loss in satellites.}
Normalized by time since infall and mean orbital pericentre, satellites lose stellar mass at a faster rate in all SIDM models considered relative to CDM. The discrepancy between CDM and SIDM grows towards lower masses in velocity-dependent models, with large differences in satellites with $M_{\rm{infall}} < 10^{9.1} \, \mathrm{M_{\sun}}$, and little difference in satellites with $M_{\rm{infall}} > 10^{9.1} \, \mathrm{M_{\sun}}$. In velocity-independent cases, the difference increases weakly with higher mass subhaloes.

\item \textbf{Subhalo evaporation does not affect MW satellites for viable SIDM models.}
Based on the mean time for collisions between a satellite particle and a host halo background particle, subhalo evaporation has virtually no effect on the mass-loss rate of satellites in v-d SIDM models. In the v-i SIDM1 model, a very small effect may exist, backed by simulations hinting at an increased mass-loss of a few percent in satellites $10 \, \mathrm{Gyrs}$ after accretion. The effect, however, is substantially less important than the precise orbit traversed by satellites. Only for the ruled out cross-section of $10 \, \mathrm{cm^2/g}$ do effects of subhalo evaporation become important.

\item \textbf{The level of enhanced stellar stripping can be predicted by the ratio \boldmath$r_{\rm{core}}/r_{\rm{t}}$.}
The relative strength of the difference in stellar mass loss between SIDM and CDM is proportional to the SIDM core radius, $r_{\rm{core}}$, divided by the tidal radius, $r_{\rm{t}}$. In v-d models, $r_{\rm{core}}$ grows much more slowly with subhalo mass than $r_{\rm{t}}$, leading to more CDM-like behaviour for larger haloes. In v-i models, $r_{\rm{core}}$ grows slightly faster than $r_{\rm{t}}$.

\item \textbf{In velocity-independent SIDM models, tidal stripping is much less efficient for low pericentre orbits than in velocity-dependent models.}
A large core forms in the host halo in v-i models, whereas only a small core forms in the v-d models. Consequently, tidal forces in v-i models are reduced for orbits with $r_{\rm{peri}} \lesssim 20 \, \mathrm{kpc}$ compared to CDM and v-d SIDM. The reduced tidal forces counteract the reduced central binding energy of stars in satellites resulting in little stellar mass-loss relative to CDM. For very low pericentre orbits, $r_{\rm{peri}} \lesssim 15 \, \mathrm{kpc}$, and large satellite cores in the v-d models, $r_{\rm{t}}$ can lie within a halo's core, reducing $r_{\rm{t}}$ relative to cuspy haloes and further accentuating the increase in stellar stripping.

\item \textbf{SIDM suppresses the stellar mass function of satellites.}
As a consequence of increased stellar mass-loss, the stellar mass function is reduced at all satellite mass scales in accordance with the ratio $r_{\rm{core}}/r_{\rm{t}}$ for some characteristic orbit.

\item \textbf{SIDM increases the half-light radius of satellites.}
Due to growing core sizes, stars tend to disperse out to larger radii in SIDM than in CDM. In addition to increasing their likelihood of being tidally stripped, the dispersal leads to larger half-light radii, which scales in accordance with the size of $r_{\rm{core}}$.

\item \textbf{SIDM does not produce easily identifiable global signatures on the stellar halo.}
The stellar halo is dominated by $\sim 10$ large accretions with $M_{\rm{infall}} > 10^{10} \, \mathrm{M_{\sun}}$. In our v-d models, these large satellites form small enough cores to leave no imprint on the stellar density profile, velocity dispersion profile, nor spatial distribution of stars on an all sky projection. In SIDM1, there are likely small imprints if all other variables remain unchanged, but the specific orbits of the largest accretions dominate any SIDM signal.

\item \textbf{Ultrafaint dwarfs may be key to observationally distinguishing between v-i SIDM, v-d SIDM, and purely baryonic core formation.}
Since the cusp/core issue and too big to fail problem mostly concern galaxies on the scale of the Milky Way classical dwarfs, leading theories of core formation all have mechanisms to resolve the central density issues on mass scales of approximately $10^{9.5} < M_{\rm{infall}} < 10^{10.5} \, \mathrm{M_{\sun}}$. Extending the theories to lower mass satellites, in particular those with $M_{\rm{infall}} < 10^{9.1} \, \mathrm{M_{\sun}}$, leads to divergent predictions among the possibilities. Finding evidence of large cores in ultrafaint dwarfs would give support to v-d SIDM. Finding small cores would be consistent with v-i SIDM, and finding cuspy density profiles would support pure CDM, and in turn strengthen the case for purely baryonic means of core formation in galaxies with sufficient star formation. By comparing half-light radii and enclosed dynamical masses of ultrafaint dwarfs to those of cored and cuspy satellites in simulations, \cite{Penarrubia10} find stronger evidence for cuspy haloes. Cored satellites lose too much mass to stripping at low galactocentric distances. Further studies incorporating more recently discovered ultrafaint dwarfs will be critical to solidify an answer.
\end{enumerate}

While our simulations and theory are sufficient to justify each of these conclusions, full hydrodynamic simulations are needed to better determine the precise magnitude of each of effect, subject to uncertainty in the strength of baryonic core formation mechanisms on top of SIDM. Additionally, even though we discuss all consequences of cores in terms of SIDM, similar concepts will apply to cores formed via other means. In particular, the migration of stars to larger radii, enhanced stellar stripping, suppression of the stellar mass function, and increase in the half-light radius would all apply generically to satellites with cores that grow gradually over time. The ratio $r_{\rm{core}}/r_{\rm{t}}$ would still be an indicator of the strength of each of these effects.

\section*{Acknowledgements}
GAD acknowledges support from an NSF Graduate Research Fellowship under Grant No. 1122374 and support from the Dr. Pliny A. and Margaret H. Price Endowment Fund as a Price Visiting Graduate Student, administered through the Center for Cosmology and AstroParticle Physics. The Dark Cosmology Centre is funded by the DNRF. JZ is supported by the EU under a Marie Curie International Incoming Fellowship, contract PIIF-GA-2013-62772.


\clearpage

\appendix
\section{Particle Tagging Verifications}
\label{sec:tagging_tests}

In Section \ref{sec:methods}, we described our particle tagging technique and the challenges arising from scattering particles in SIDM simulations. In Fig.~\ref{fig:tag_distr_grid}, we presented one test of our methodology, showing how particles were selected in similar fashion according to their boundedness percentile in each SIDM-CDM pair. In Section~\ref{sec:stellar_halo}, we verified that the slope of the stellar density profile in the outer halo matches that of a hydro simulation. Here, we list the additional tests taken to ensure our results do not arise from tagging artefacts.

\begin{enumerate}

\item
We checked the distance to the median tagged particle and the furthest tagged particle versus infall mass for each CDM and SIDM simulation pair. Since the central density of haloes in SIDM is reduced relative to haloes in CDM, there are fewer particles within a fixed radius. This raises the concern that while particles are tagged in the same distribution in terms of rank order boundedness, particles in SIDM haloes are tagged at systematically further radii. Thus, enhanced stripping could be due to star particles starting at larger radii and nothing else. Per infall mass and infall time bin, we find SIDM particles are either tagged at no further distance than CDM particles or are tagged at fractionally larger radii. These are small differences in all cases, which are negligible compared to the growth in $r_{1/2}$ in SIDM relative to CDM seen in Fig.~\ref{fig:half_mass_radius}.

\item We tested if haloes with large amounts of internal scattering, and thus very few particles to tag, produce biased results. We isolated all infalling haloes where we could tag $<100\%$, $<50\%$, and $<20\%$ of our target of $2\%$ of the most bound particles. Hereafter, we call the fraction of the target number particles to tag that can actually be tagged the tagging yield. When removing each of these subsets, all trends of enhanced stellar stripping as seen in Figs~\ref{fig:strip_v_time_p2}-\ref{fig:strip_v_time_p4} remain the same.

\item We further tested if haloes with very few particles to tag, either due to their small size or large fraction of scattered particles produce biased results. We therefore removed all haloes with $<N$ particles tagged from our data, with $N$ ranging from $5$ to $10$, and again find the trends in enhanced stellar stripping unchanged.

\item We tested for bias due to ranges of infall time and infall mass where many SIDM haloes have zero particles to tag. In our tagging process, we divide haloes into bins by infall time and infall mass, and use the mean distribution of what particles can be tagged in each bin in the SIDM case to apply to the corresponding CDM case. As a result, the group of CDM haloes in each bin are all tagged with the same distribution, whereas the SIDM haloes are tagged with some variation about that distribution. In bins where the typical tagging yield is below $20\%$, a fraction of the SIDM haloes have no particles to tag, and are thus not included in further analysis, whereas the CDM haloes are all tagged with the same low tagging yield. Concerned about this mismatch in the number of haloes included in our two samples, we investigated the trends in enhanced stellar stripping after excluding bins with substantial mismatches. Once again, this did not change the trends.

\item
We searched for biases in haloes with early and late infall times. For a fixed infall halo mass, haloes with early infall times have more internal scattering between infall and $z=0$ than those with late infall times. The early infall haloes have particles typically tagged at larger distances, and in turn the fraction of particles stripped is consistently higher. However, the difference in the strength of stripping between SIDM and CDM remains the same for both early and late infall time samples.
\end{enumerate}

\bsp	
\label{lastpage}
\end{document}